\documentclass[prb, twocolumn, aps,amsmath,amssymb,amsfonts, superscriptaddress]{revtex4}
\usepackage[german,american,english]{babel}
\usepackage{graphicx}
\usepackage{graphics}
\usepackage{dcolumn}
\usepackage{multirow}
\usepackage{bm}
\usepackage{latexsym}
\usepackage{amssymb}
\usepackage{amsmath}
\usepackage{amsfonts}
\usepackage{layout}
\usepackage{verbatim}
\usepackage{epsfig}
\usepackage{graphicx}
\usepackage{amsbsy}
\usepackage{xcolor}
\usepackage{enumitem}
\usepackage[caption=false]{subfig}
\newcommand{\bea}{\begin{eqnarray*}}
	\newcommand{\eea}{\end{eqnarray*}}
\newcommand{\bne}{\begin{equation*}}
\newcommand{\ede}{\end{equation*}}

\newcommand{\bnen}{\begin{equation}}
\newcommand{\eden}{\end{equation}}
\newcommand{\bean}{\begin{eqnarray}}
\newcommand{\eean}{\end{eqnarray}}
\newcommand{\bsen}{\begin{subequations}}
	\newcommand{\esen}{\end{subequations}}

\newcommand{\bna}{\begin{array}}
	\newcommand{\eda}{\end{array}}
\newcommand{\bnm}{\begin{enumerate}}
	\newcommand{\edm}{\end{enumerate}}

\newcommand {\bra} [1] {\langle #1 |}
\newcommand {\ket} [1] {| #1 \rangle}

\newcommand {\tbkt} [3] {\langle #1 | #2 | #3 \rangle}

\begin{document}
	
	\title{Optimisation of electrically-driven multi-donor quantum dot qubits}
	
	\author{Abhikbrata Sarkar}
	\affiliation{School of Physics, The University of New South Wales, Sydney 2052, Australia}
	\affiliation{ARC Centre of Excellence in Future Low-Energy Electronics Technologies, The University of New South Wales, Sydney 2052, Australia}
	\author{Joel Hochstetter}
	\affiliation{School of Physics, The University of New South Wales, Sydney 2052, Australia}
	\author{Allen Kha}
	\affiliation{School of Physics, The University of New South Wales, Sydney 2052, Australia}
	\author{Xuedong Hu}
	\affiliation{Department of Physics, University at Buffalo, SUNY, Buffalo, NY 14260-1500}
	\author{Michelle Y. Simmons}
	\affiliation{Centre for Quantum Computation and Communication Technology, School of Physics,
		The University of New South Wales, Sydney, New South Wales 2052, Australia}
	\author{Rajib Rahman}
	\affiliation{School of Physics, The University of New South Wales, Sydney 2052, Australia}
	\author{Dimitrie Culcer}
	\affiliation{School of Physics, The University of New South Wales, Sydney 2052, Australia}
	\affiliation{ARC Centre of Excellence in Future Low-Energy Electronics Technologies, The University of New South Wales, Sydney 2052, Australia}
	
	\begin{abstract}
		Multi-donor quantum dots have been at the forefront of recent progress in Si-based quantum computation. Among them, $2P:1P$ qubits have a built-in dipole moment, making them ideal for electron dipole spin resonance (EDSR) using the donor hyperfine interaction, and thus all-electrical spin operation. The development of all-electrical qubits requires a full understanding of their EDSR and coherence properties, in which multi-donor dot qubits are expensive to model computationally due to the multi-valley nature of their ground state. Here, by introducing a variational effective mass wave-function, we examine the impact of qubit geometry and nearby charge defects on the electrical operation and coherence of $2P:1P$ qubits. We report four outcomes: (i) The difference in the hyperfine interaction between the $2P$ and $1P$ sites enables fast EDSR, with $T_\pi \sim 10-50\ $ns and a Rabi ratio $ (T_1/T_\pi) \sim 10^6$. We analyse qubits with the $2P:1P$ axis aligned along the [100], [110] and [111] crystal axes, finding that the fastest EDSR time $T_\pi$ occurs when the $2P:1P$ axis is $\parallel$ [111], while the best Rabi ratio occurs when it is $\parallel$ [100]. This difference is attributed to the difference in the wave function overlap between $2P$ and $1P$ for different geometries. In contrast, the choice of $2P$ axis has no visible impact on qubit operation. (ii) Sensitivity to random telegraph noise due to nearby charge defects depends strongly on the location of the nearby defects with respect to the qubit. For certain orientations of defects random telegraph noise has an appreciable effect both on detuning and $2P-1P$ tunneling, with the latter inducing gate errors. (iii) The qubit is robust against $1/f$ noise provided it is operated away from the charge anticrossing. (iv) Entanglement via exchange is several orders of magnitude faster than dipole-dipole coupling. These findings pave the way towards fast, low-power, coherent and scalable donor dot-based quantum computing.
	\end{abstract}
	\date{\today}
	\maketitle
	\section{Introduction}\label{sec:1}
Quantum computation architectures require long coherence times and a clear route towards scaling up. \cite{Koiller2002,Awschalom2007} Solid state spin qubits \cite{loss1998quantum,levy2002universal,kane1998silicon, petta2005coherent} are excellent candidates for large scale quantum computation \cite{hill2005global} with outstanding coherence and fidelity, with Si:P donors \cite{kane1998silicon,morello2010single,buch2013spin,o2001towards} having a number of advantageous features. The strong Coulomb confinement potential of the donor atom comes for free and is reproducible, which, when coupled with extensive materials knowledge from the Si microfabrication industry, constitutes a viable avenue towards scalability. Thanks to the weak spin-orbit coupling\cite{weber2018spin} of Si electrons, the presence of hyperfine-free isotopes,\cite{tahan2005rashba,tyryshkin2012electron} and absence of piezoelectric coupling to phonons, \cite{prada2008singlet,zwanenburg2013silicon} the coherence time of Si:P donor electron spins is the longest among solid-state qubits. \cite{feher1959electron1, feher1959electron2, tyryshkin2003electron} Exceptional experimental progress has been registered in the past decade. \cite{weber2014spin,he2019two, kuhlmann2013charge,
		weber2018spin,		tyryshkin2003electron,	watson2015high, laird2007hyperfine,	wang2017all,
		Veldhorst2015two, morello2010single,
		muhonen2015quantifying, shamim2011suppression}





	


Optimisation of the speed and scalability of multi-donor quantum dot qubits will improve with our ability to operate them using only electric fields. Electric fields are much easier to apply and localise than magnetic fields, can be switched much faster, and consume less power. Spin qubits in single donors have exceptionally long coherence times but, lacking an intrinsic dipole moment, are challenging to operate electrically. Moreover, the multi-valley nature of the ground state (GS) \cite{ning1971multivalley, rahman2009orbital, gamble2015multivalley} makes the exchange coupling \cite{koiller2001exchange, koiller2004shallow} dependent on individual donor positions within the crystal. These oscillations have been mitigated by placing the donors along certain crystallographic directions \cite{voisin2020valley} or by the presence of strain, \cite{wellard2003electron} while the use of multi-donor quantum dots provides another approach to reduce this atomistic scale sensitivity by introducing valley-weight anisotropy. \cite{hollenberg2006two, wang2016highly, voisin2020valley} Multi-donor quantum dot qubits can also be designed with a built-in dipole moment by making the charge densities different in adjacent dots, which enables electron dipole spin resonance (EDSR).\cite{wang2017all, krauth2021flopping, osika2021spin} At the same time, dipole moments are known to expose the qubit to charge noise. The main questions at this stage in the development of multi-donor quantum dot qubits are: \textit{What is the largest achievable EDSR Rabi ratio in a donor-dot qubit, and can such an electrically operated qubit be robust against charge noise?} These questions are theoretically challenging, since multi-donor dots are more difficult to model analytically and very expensive to treat computationally. \cite{rahman2011stark, klymenko2015multivalley, saraiva2015theory} In addition, an understanding of EDSR and coherence properties in these multi-donor systems requires one to treat spatially and temporally random functions, which further increase the complexity and computational cost of the problem.
	
	
With these observations in mind, in this work we develop a variational effective mass wave function (EMA) wave function for $2P$ \cite{watson2015high, buch2015quantum} and $2P:1P$\cite{wang2016highly} multi donor quantum dots in Si, and use it to study quantitatively the properties of a $2P:1P$ double donor quantum dot spin qubit. It was recently shown that strong EDSR can be achieved in donor quantum dots using the hyperfine interaction, which plays the role of an effective spin-orbit field built into the qubit.\cite{osika2021spin, krauth2021flopping} We focus on the following properties: (i) the impact of qubit geometry on the EDSR gate time and Rabi ratio, and (ii) the impact of nearby charge defects, inducing random telegraph noise, on the coherence properties of the $2P:1P$ spin qubit. 

To determine the role of geometry in the electrical operation of the qubit, we consider three different orientations of the 2P:1P axis (the \textit{qubit axis}): $[100]$, $[110]$ and $[111]$, and compare the time scales relevant to qubit operation for these orientations. We find that the fastest EDSR time $T_\pi$ is achieved when the qubit axis is $\parallel$ [111], while the largest Rabi ratio is found when the qubit axis is $\parallel$ [100]. This can be explained by the difference in the wave function overlap between the $2P$ and $1P$ sites: both $T_\pi$ and $T_1/T_\pi$ decrease linearly with $t$, while the tunneling $t$ is highest for the $2P:1P$ axis $\parallel$ [100]. This is also the direction along which the $2P$ and $1P$ wave functions have the largest overlap, resulting in the slowest $T_\pi$ and largest Rabi Ratio. The tunnel coupling $t$ is smallest for [111], hence $T_\pi$ is the fastest. In contrast, the orientation of the $2P$ axis only changes $T_\pi$ by 1-2$\%$, and we conclude it has no visible effect on qubit operation. Since aligning the qubit axis $\parallel$ [100] yields the largest Rabi ratio, in the latter part of the paper, devoted to coherence, we focus on this particular geometry. For the spin relaxation time $T_1$, our theory yields $1/T_1 \propto B^5$, consistent with an acoustic phonon mediated valley population mechanism.\cite{hasegawa1960spin} This result in the low temperature limit is in accordance with earlier single-shot readout experiments.\cite{hanson2007spins, morello2010single, watson2017atomically} 

Next we provide a quantitative analysis of the effect of charge defects on qubit operation. Importantly, a nearby charge defect can have a significant effect on both the detuning and $2P:1P$ tunnelling, an aspect for which no quantitative studies exist in donor systems, although the issue has been known to exist in quantum dot qubits. \cite{hu2006charge, culcer2009dephasing}
The contributions of the defect potential to detuning and tunnelling depend strongly on the defect location and orientation with respect to the qubit axis. As a result of the change in the $2P:1P$ tunnel coupling, random telegraph noise can affect the EDSR gate fidelity in the vicinity of the anti-crossing. On the other hand, for $1/f$ noise the effect on the $2P:1P$ tunnelling due to an ensemble of nearby defects is expected to be washed out, and operating the qubit away from the anti-crossing will drastically reduce its sensitivity to $1/f$ noise. In addition, our calculation shows that the anisotropic hyperfine interaction due to dipolar coupling between the electron and nuclear spins causes negligible decoherence. Finally, we examine entanglement via exchange\cite{loss1998quantum,kane1998silicon} and dipole-dipole coupling\cite{trif2007spin,flindt2006spin} and find the former to be several orders of magnitude more efficient than the latter. Taken together, these findings pave the way towards high-performance all-electrical, coherent and scalable quantum computation using multi-donor quantum dots.
	
The outline of this paper is as follows. In section \ref{sec:2}, we present the variational wave function for a $2P$ donor-dot. Section \ref{sec:3} outlines the $2P:1P$ qubit architecture, all-electrical qubit control, fidelity and coherence by studying EDSR, phonon relaxation and dephasing, and also explores the long-distance qubit couplings. We end with a brief summary.
	 
\section{Variational $2P$ wave function} \label{sec:2}
	
The effective Hamiltonian for one electron in a double donor quantum dot is written as \cite{pantelides1978electronic,kittel1954theory}
\begin{equation}\label{eq:donor2DHamil}
	H_{2d} = -\frac{\hbar^2\nabla^2}{2m^*} + \!\! \sum_{D = L, R} \bigg(-\frac{e^2}{4\pi\varepsilon_0\varepsilon_r|{\bm r} - {\bm R}_D|} + H^v_D\bigg)
	\end{equation}
	where $m^*$ denotes the effective mass, including anisotropy within the quantum dot (the longitudinal effective mass $m_l$=0.916$m_e$ and the  transverse effective mass $m_t$=0.191$m_e$, with $m_e$ the bare electron mass), ${\bm R}_L$ and ${\bm R}_R$ stand for the position of the left and right donors in the $2P$ donor quantum dot respectively, and $H^v_D$ represents the short-range part of the Coulomb potential of each donor, which gives rise to the valley-orbit coupling, discussed below. 
	
	\begin{figure*}[!htbp]
	\subfloat[]{\begin{minipage}[t]{0.33\linewidth}
\includegraphics[width=\textwidth]{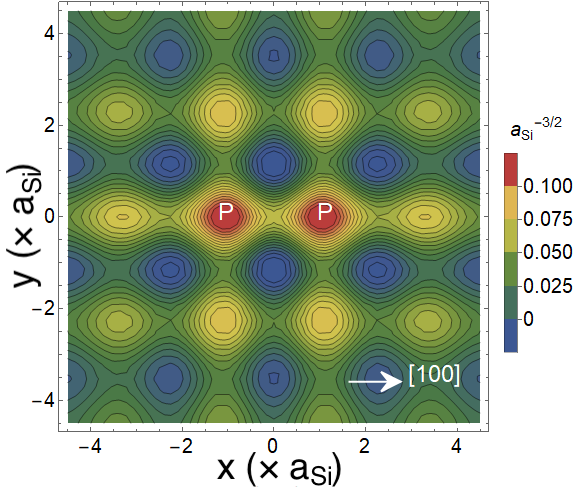}
\label{fig:2dwf100}
\end{minipage}}
\hfill
\subfloat[]{\begin{minipage}[t]{0.33\linewidth}
\includegraphics[width=\textwidth]{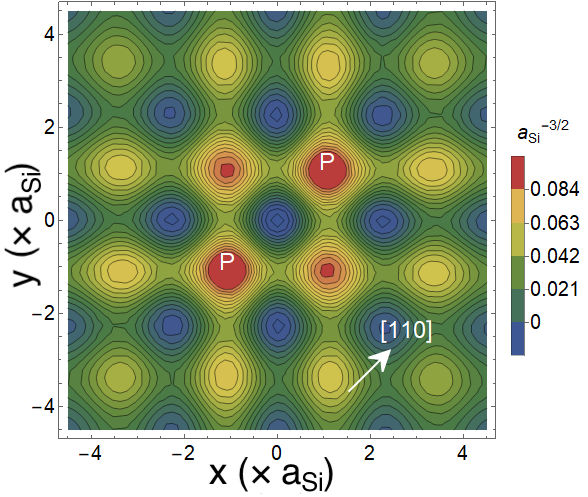}
\label{fig:2dwf110}
\end{minipage}}
\hfill
\subfloat[]{\begin{minipage}[t]{0.33\linewidth}
\includegraphics[width=\textwidth]{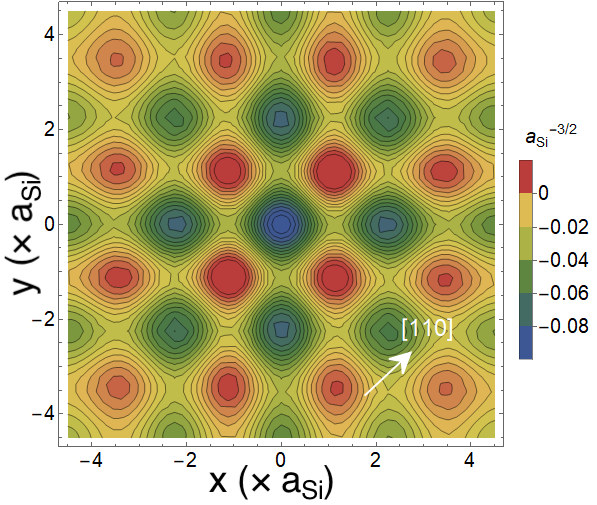}
\label{fig:2dwf111}
\end{minipage}}
	\caption{\textbf{Spatial dependence of the variational $2P$ wave function amplitude in the effective mass approximation}. The length scale is the lattice constant of Si ($a_{Si}$), with Bloch function $u(\mathbf{r})=1$. a) The principal maxima are at the two donors' positions: $(-1,0)$ and $(1,0)$ in the $x$-$y$ plane, with secondary maximum due to short-range traveling waves at $(-1,-2)$,$(1,2)$ and so on. The overall diminishing amplitude away from the donors is determined by the envelope function. Here the $2P$ multi donor quantum dot is oriented along $[100]$. b) Two donors' positions: $(-1,-1)$ and $(1,1)$ in the $x$-$y$ plane, with secondary maximum at $(-1,1)$,$(1,-1)$ and so on. The $2P$ multi donor quantum dot is oriented along $[110]$. c) in-plane amplitude variation of $2P\parallel[111]$ with two donors' positions: $(-1,-1,-1)$ and $(1,1,1)$. Only diminished secondary maximum in $x$-$y$ plane are captured.}
	\end{figure*}
	
	The EMA wave functions for individual donors at position ${\bm r}$ are given by \cite{Kohn1955,feher1959electron1} $D_\xi ({\bm r}) = \phi_{D,\xi}({\bm r} - {\bm R}_D) \, e^{i{\bm k}_\xi \cdot({\bm r} - {\bm R}_D)} \, u_\xi ({\bm r}),$ where $\xi$ is the index for the six valleys in Si, and $|{\bm k}_\xi| = \pm k_0 \equiv \pm 0.85 \, (2\pi/a_{\rm Si})$, with $a_{\rm Si} = 5.43\AA$ the lattice constant of Si. The lattice-periodic function is $ u_\xi ({\bm r}) = \sum_{\bm K} c^\xi_{\bm K} e^{i{\bm K}\cdot{\bm r}}$, with ${\bm K}$ reciprocal lattice vectors. Here $\phi_{D,\xi}({\bm r}-{\bm R}_D)$ denotes the hydrogenic part of the wave function. The resultant states\cite{koster1954wave} contributing to the envelope of the donor wave function\cite{salfi2014spatially} take the form of a deformed hydrogenic $1$s orbital with two radii $a$ and $b$,
	\begin{equation}\label{eqn:envelop}
	\phi_{\pm x} ({\bm r}) = \frac{1}{ \sqrt{\pi a^2b}} \, e^{-\sqrt{\frac{x^2}{b^2}+\frac{y^2+z^2}{a^2}}}
	\end{equation}
	$\phi_{\pm y} ({\bm r})$ and $\phi_{\pm z} ({\bm r})$ can be obtained from Eqn.~\ref{eqn:envelop} by interchanging $y\leftrightarrow x$ and $z\leftrightarrow x$ repectively.
	The single-donor variational wave function is given by: $\Psi_{1D}(\mathbf{r})=\frac{1}{\sqrt{6}} \sum_{\xi}\phi_{D,\xi}(\mathbf{r}) e^{i\mathbf{k_\xi\cdot \mathbf{r}}}u_\xi(\mathbf{r})$. 
	
	The valley-orbit coupling\cite{baldereschi1970valley, friesen2010theory, saraiva2009physical, culcer2010interface, calderon2007external} is added to a single donor through the term $H^v_D = U_0 \, \delta ({\bm r} - {\bm R}_D)$. In the basis spanned by the six valley wave functions $\{ \phi_{D,\xi}(\mathbf{r}) e^{i\mathbf{k_\xi\cdot \mathbf{r}}}u_\xi(\mathbf{r})\}$ this will yield the standard Kohn-Luttinger matrix elements $\Delta_\parallel$, $\Delta_\perp$ and $\delta$, and we choose the value of $U_0$ to reproduce the correct energy splitting between the ground and first excited states. \cite{ramdas1981spectroscopy} This enables us to circumvent complications associated with the central cell correction $r_{cc}$: a single $r_{cc}$ cannot be chosen to produce both the donor binding energy and the valley-orbit coupling correctly.\cite{gamble2015multivalley} 	

\begin{table}
\begin{tabular}{ r |c| c| c| c| c } 
 \hline
 \hline
 Parameter & \multicolumn{3}{c|}{EMA} & TB & FEM \\
 \cline{2-6}
 & $[100]$ & $[110]$ & $[111]$ & $[100]$ & [100]\\
 \hline
 Longitudinal Bohr radius,& & &  & &\\
  $a (nm)$ & $1.322$ &$1.357$ & $1.385$  & $1.4$ &\\
  \hline
 Transverse Bohr radius,& & &  & \\
  $b (nm)$ & $0.736$  & $0.763$ & $0.784$ & $0.73$ &\\
  \hline
$2P$ bare GS energy, & & &  & \\ 
$E_{2P}(meV)$ & $-105$ & $-102$ & $-99$ &  N/A &\\ 
\hline
 Valley-orbit correction,& & &  & \\
  $\delta (meV)$ & $-16$ & $-14$ & $-12$ & N/A &\\ 
  \hline
 $2P$ corrected GS energy, & & &  & \\
 $E_{2P}^0(meV)$ & $-121$ & $-116$ & $-111$ & $-130$ & $-120$\\
 \hline
 \hline
\end{tabular}
\caption{Comparison of key metrics of 2P double donor dots using the variational method effective mass approximation (EMA) with the $2P$ axis aligned with $[100]$, $[110]$ and $[111]$. The tight-binding (TB) values are taken from Refs.~\onlinecite{voisin2020valley,weber2014spin}. The finite-element method (FEM) value in col. 4 is taken from Ref.~\onlinecite{klymenko2014electronic}.}
\label{tab:variarional}
\end{table}
For a 2P quantum dot, the wave functions for the left and right donors are given by $\phi_{L,\xi} ({\bm r })=\phi_{\xi} (\bm{ r-R_L})$ and $\phi_{R,\xi} (\bm{ r })=\phi_{\xi} (\bm{ r-R_R})$ respectively. We determine the dot wave function based on the variational method developed in Ref.~\onlinecite{slater1974self} for the hydrogen molecule, then add the valley-orbit coupling due to the two donors as above. The variational method has been shown to reproduce the spectra of H$_2$ and He with great accuracy. \cite{slater1974self} We treat $a$ and $b$ as variational parameters\cite{slater1974self} which are a function of donor distance, and define $\ket{F_{D,\xi}} = \phi_{D,\xi} \, e^{i\mathbf{k_\xi\cdot({\bm r} - {\bm R}_D)}}$.   
The matrix elements of the Hamiltonian $H_{\pm \xi}=\begin{pmatrix}
\varepsilon_\xi & \tilde{t}_\xi \cr
\tilde{t}_\xi & \varepsilon_\xi
\end{pmatrix}$ between the wave functions for the same valley $\{ F_{L,\xi}, F_{R,\xi}\}$ are the on-site energies $\varepsilon_\xi=\bra{\phi_{D,\xi}}H_{2d}\ket{\phi_{D,\xi}}$ and the tunneling energies $\tilde{t}_\xi=\bra{\phi_{L/R,\xi}}H_{2d}\ket{\phi_{R/L,\xi}}$, with $\varepsilon_\xi, \tilde{t}_\xi$ real. These matrices are diagonalised by symmetric and anti-symmetric combinations $\ket{S_{\xi}}$, $\ket{A_{\xi}}$, which provide an orthonormal basis of valley wave functions. Since the $\tilde{t}_\xi$'s are negative, the symmetric functions $\ket{S_\xi}$ have a lower eigen-energy $E_\xi = \varepsilon_\xi - |\tilde{t}_\xi|$ than the anti-symmetric functions $\ket{A_\xi}$ with $E_\xi = \varepsilon_\xi + |\tilde{t}_\xi|$. We can ignore the anti-symmetric wave function as the inter-valley matrix elements are small compared to the energy difference of $2\tilde{t}_\xi$ between the symmetric and anti-symmetric functions. We are left with a $6\times 6$ matrix in the manifold spanned by $\ket{S_{\xi}}$. Minimization of $\tbkt{S_\xi}{H}{S_\xi}$ yields the variational parameters $a$ and $b$ (Table~\ref{tab:variarional}). For $2P\parallel [100]$ the ground state energy is $-105$ meV and the valley splitting is $-16$ meV. Hence, the valley-orbit corrected ground state energy is $-121$ meV, which compares very well with the value of $-120$ meV using finite-element method (FEM) in Ref.~\onlinecite{klymenko2014electronic} and effective mass approximation (EMA) in Ref.~\onlinecite{saraiva2015theory}, especially considering the simplicity of our method. For $[110]$ and $[111]$, the valley-orbit corrected $2P$ ground state energy is $-116$ meV and $-111$ meV respectively.

The simple model we have devised for the $2P$ wave function is exactly diagonalisable analytically and can treat arbitrary donor position and donor dot separation and orientation easily, avoiding complications associated with central cell corrections to the valley-orbit coupling, outlined below. \cite{fetterman1971field, oliveira1986effect} Comparison of the ground state energy, valley composition and exchange oscillations between our method and much more advanced computational approaches such as finite-element method \cite{klymenko2014electronic}, Nano-Electronic MOdelling\cite{klimeck2007atomistic} is surprisingly encouraging.

Figure \ref{fig:2dwf100} depicts the color coded amplitude of the $2P$ quantum dot wavefunction\cite{klymenko2014electronic} in the $x$-$y$ plane with the single donors situated at the closest lattice spacing of $a_{Si}=\,0.54\,nm$ along the $[100]$ direction, using the envelope function and incorporating  the phase factor $e^{i {\textbf K} \cdot {\textbf r}}$ that determines the short-range structure of $\Psi_{2D}$. In contrast the $2P\parallel[110]$ wavefunction is shown in Fig.~\ref{fig:2dwf110}, where the closest spacing between the two donors is $\sqrt{2}a_{Si}=0.76$ nm. Fig.~\ref{fig:2dwf111} shows the diminished secondary maximum of the $2P\parallel[111]$ wavefunction in the $x$-$y$ plane, with both donors out-of-plane. The ground-state wave function simplifies to $\Psi_{2D}(\mathbf{r})=\frac{1}{\sqrt{\sum_{\xi}w_{\xi}^2}} \sum_{\xi}  w_{\xi} S_\xi u_\xi(\mathbf{r})$, with $w_\xi$ representing the valley weight. The ground state of the double-donor quantum dot is symmetric under inversion yet, unlike the case of a single-donor quantum dot, the wave function is no longer spherically symmetric. Importantly this spatial anisotropy has a significant effect on the valley composition of the ground state. The large effective mass anisotropy (EMA) results in a large enhancement of the kinetic energies of the valley states lying perpendicular to the orientation of the $2P$ quantum dot. When the $2P$ axis is $\parallel [100]$ the contribution to the kinetic energy stemming from the $x$-valleys will be smaller by a factor of $4.8$ than that due to the $y$ and $z$ valleys perpendicular to the $2P$ axis. In this case the ground state would comprise only the $y$ and $z$ valleys. The valley-orbit coupling (VOC) arising from short-range Coulomb potential $H^v_D$ (Eqn.~\ref{eq:donor2DHamil}) on the other hand is responsible for matrix elements of approximately the same magnitude connecting all valleys, and therefore favours a ground-state superposition of all valleys with comparable weight. The competition between the effects of effective mass anisotropy and valley-orbit coupling ultimately determine the ground state valley weight ratio.
\begin{itemize}[noitemsep,nolistsep]
    \item When the $2P$ axis $\parallel [100]$ the valley weight ratio $w_{\pm x}:w_{\pm y}:w_{\pm z}= 0.88:1:1$, indicating a higher contribution to $\Psi_{2D}$ from the $y$ and $z$ valley states.
    \item When $2P$ axis is $\parallel [110]$ the valley weight ratio $w_{\pm x}:w_{\pm y}:w_{\pm z}= 0.91:0.91:1$ indicating equal and lower contribution from $x$ and $y$ valleys.
    \item When the $2P$ axis is $\parallel [111]$ all valley weights are equal, as the effects of EMA and VOC are symmetric in all valleys.
\end{itemize}  
To summarise this section, we have presented a simple and practical variational wave function describing the ground state of a $2P$ donor dot, whose structure reflects the interplay between the effective mass anisotropy and valley-orbit coupling, to yield an accurate value for the ground state energy. This wave function is very convenient in formulating a simple analytical model for the operations of a $2P:1P$ qubit, as the next section shows. 

\section{$2P:1P$ Qubit} \label{sec:3} 

In order to operate a multi-donor dot qubit electrically we require a dipole moment, which a $2P$ donor dot does not have. On the other hand, a double donor dot qubit in a 2P:1P configuration has a difference in the charge densities between the $2P$ and $1P$ sites which gives rise to a dipole moment that can couple to an applied electric field. 
In what follows we concentrate on such a $2P:1P$ qubit and discuss the electrical operation and EDSR using the hyperfine coupling to the nuclei. Next, considering the fact that this dipole moment also couples to phonons and charge fluctuations, we discuss the coherence properties of an electrically operated $2P:1P$ qubit. Finally, we discuss briefly the prospects for entangling two $2P:1P$ qubits using the exchange interaction as well as the dipole-dipole interaction. 

\subsection{Effective Hamiltonian for a $2P:1P$ qubit} 

\begin{figure}[tbp]
\subfloat[]{\begin{minipage}[c]{0.5\linewidth}
\includegraphics[width=1.7 in, height= 1 in]{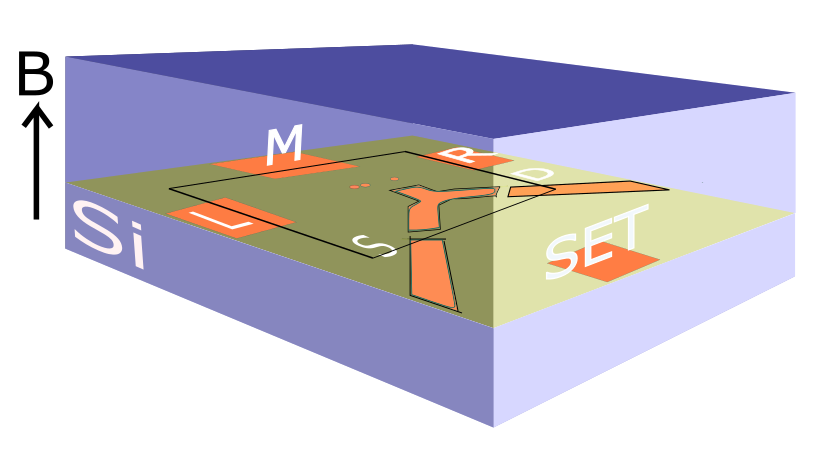}
\label{fig:schematica}
\end{minipage}}
\hfill
\subfloat[]{\begin{minipage}[c]{0.5\linewidth}
\includegraphics[width=1.6 in, height= 1.2 in]{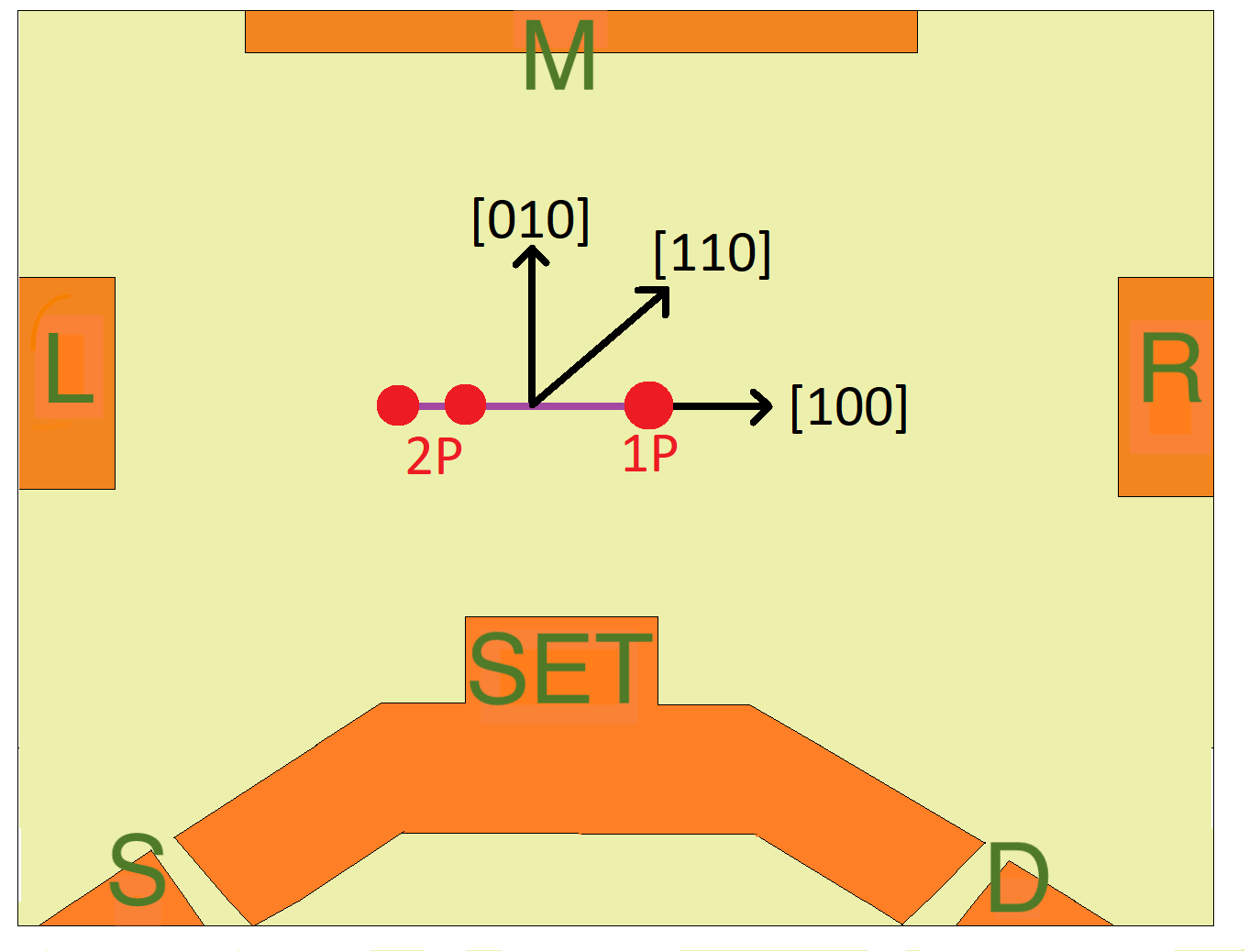}
\label{fig:schematicb}
\end{minipage}}
\caption{\textbf{Schematic of a $2P:1P$ EDSR qubit architecture, illustrating the way our qubit design can be adapted to atomic scale lithographic techniques using in-plane gates}. a) The Si matrix (blue box) is shown, with the highlighted green plane containing the physical $2P:1P$ qubit along the $[100]$ direction, as emphasised on the right. A schematic of the STM image of the proposed qubit device is presented following Ref.~\onlinecite{he2019two}; M=middle gate, L=left gate, R=right gate, S=source, D=drain, and SET signifying single electron transistor gate. b) A qubit geometry with both dots 2P-1P aligned along $[100]$ crystallographic direction. The left, right and middle gates mediate the charge distribution between the dots, hence control single qubit operations.} 
\end{figure}

 We first introduce the general formalism describing a $2P:1P$ qubit. We adopt the Hund-Mulliken approach, which takes into account the overlap of the $2P$ and $1P$ wave functions. Fig.~\ref{fig:schematica} represents a schematic of the STM device architecture showing the proposed qubit system, in which the $2P$ axis and the $2P:1P$ axis are both $\parallel$ [100]. The $2P$ dot is situated on the left and the $1P$ single donor quantum dot is on the right. The ground states are orthonormalised as: $\Psi_{+/-} = \left(\Psi_{1D/2D}(\mathbf{r}-\mathbf{R}_{1D/2D}) - g \Psi_{2D/1D}(\mathbf{r}-\mathbf{R}_{2D/1D})\right)/s'$,
with $s'=\sqrt{1-2gS+S^2}$; where $S$ is the overlap integral $\int d^3r\, \Psi_{2D}(\mathbf{r}-\mathbf{R}_{2D})\,  \Psi_{1D}(\mathbf{r}-\mathbf{R}_{1D})$, used for normalization, and $g=\left(1- \sqrt{1- S^2}\right)/S$. In the singlet-triplet particle state space\cite{burkard1999coupled} with the basis $\Psi_{s\mp}(\mathbf{r}_1,\mathbf{r}_2)=\Psi_\mp(\mathbf{r}_1) \Psi_\mp(\mathbf{r}_2)$, $\Psi_{s0}(\mathbf{r}_1,\mathbf{r}_2)=(\Psi_+(\mathbf{r}_1) \Psi_-(\mathbf{r}_2)+\Psi_-(\mathbf{r}_1) \Psi_+(\mathbf{r}_2))/\sqrt{2}$, $\Psi_{t}(\mathbf{r}_1,\mathbf{r}_2)=(\Psi_+(\mathbf{r}_1) \Psi_-(\mathbf{r}_2)-\Psi_-(\mathbf{r}_1) \Psi_+(\mathbf{r}_2))/\sqrt{2}$, we include the on-site energies, inter-dot tunnelling, and Coulomb interaction effects as described in Ref.~\onlinecite{culcer2010quantum}. The $2P:1P$ exchange energy is given by the difference between the two lowest eigenvalues of the $4\times4$ matrix spanned by the two-particle basis mentioned above (See supplementary material).

A detuning dc electric field is applied between the in-plane gates $M$, $L$ and $R$ to drive the $2P:1P$ qubit to the charge anti-crossing. The detuning, denoted by $\delta$, represents the corresponding energy difference between the $2P$ and $1P$ sites. The resultant ground and excited orbital state energies are $e_1=\frac{1}{2}(e_l+e_r+\delta-\delta\varepsilon)$,  $\,e_2=\frac{1}{2}(e_l+e_r+\delta+\delta\varepsilon)$. The anti-crossing occurs at the point where $e_2-e_1=\delta\varepsilon=\sqrt{(e_l-e_r+\delta)^2+4t^2}\rightarrow 2t$; $e_l$, $e_r$ being the on-site energies of $2P$ (left) and $1P$ (right), $t$ signifies the tunneling energy. The hyperfine interaction between the electron and effective nuclear field from the three donors is
  \begin{equation}\label{eqn:hyperfine} 
  H_{hf}=\mathcal{A}_0 \sum_i \delta(\mathbf{r}-\mathbf{R}_i) \overrightarrow{I_i}\cdot \overrightarrow{S}
  \end{equation}
with $\mathcal{A}_0=(2/3)\mu_0\gamma_e \gamma_n$; $\gamma_e,\gamma_n$ being the gyromagnetic ratios of the electron and proton respectively; $\overrightarrow{I}_i$ and $\overrightarrow{S}$ denote nuclear and electron spin operators and $i=1P, 2P_L, 2P_R$ label the locations of the nuclear spins.\cite{kohn1955hyperfine} We assume isotopically purified $^{28}$Si so that there are no hyperfine fields other than those due to the donors given above. An external magnetic field $B$ applied along $z$ resolves the orbital states by Zeeman energy $\pm e_z$ to produce electron spin-up and spin-down states.  \cite{krauth2021flopping,keith2021microsecond} We include a driving ac electric field $\widetilde{E}(t)$ applied between $M$, $L$ and $R$ (Fig.~\ref{fig:schematicb}), which gives rise to an oscillating electrical potential $v_{ac}(t) = e\widetilde{E}(t) x$. The total Hamiltonian $H_q$ in the $2P:1P$ orbital+spin basis $\{\overline{G}\uparrow$, $\overline{G}\downarrow$,$ \overline{E}\uparrow$, $\overline{E}\downarrow\}$ reads:

\begin{small}
			\begin{equation}\label{eqn:masterhamiltonian}
			H_q = \begin{pmatrix}
			e_1+\epsilon_z&k_{gg}b_- &|&  k_{ge}b_z+ v_{ac} & k_{ge}b_- \cr
			k_{gg}b_+ &e_1-\epsilon_z &|& k_{ge}b_+ & v_{ac}-k_{ge}b_z \cr
			-----&-----&|&-----&-----&\cr
			k_{eg}b_z+v_{ac} & k_{eg}b_- &|& e_2+\epsilon_z & k_{ee}b_- \cr
			k_{eg}b_+ & v_{ac}-k_{eg}b_z &|&k_{ee}b_+ &e_2-\epsilon_z \cr
			\end{pmatrix}
			\end{equation}
\end{small}
    
The matrix elements $k_{gg}\,{=}\,\mathcal{A}_0\tbkt{\overline{G}}{\delta(\mathbf{r})}{\overline{G}}$, $k_{ge}\,{=}\,k_{eg}\,{=}\,\mathcal{A}_0\tbkt{\overline{G}}{\delta(\mathbf{r})}{\overline{E}}$ arise from the contact hyperfine interaction (Eqn.~\ref{eqn:hyperfine}); $b_+\,{=}\,\langle I_+ \rangle,\,b_-\,{=}\,\langle I_-\rangle,\,b_z\,{=}\,\langle I_z\rangle$ are the expectation values of the nuclear spin angular momentum operators. $k_{gg}b_z$ stems from the out-of-plane hyperfine interaction, and this is simply added to the total Zeeman splitting $e_z$, modifying the spin states splitting as $\epsilon_z\,{=}\,e_z+k_{gg}b_z$. In summary, we have determined a simple, exactly diagonalisable analytical Hamiltonian that describes the electrical operation of a $2P:1P$ qubit. This will be the topic of the next section.

\begin{figure*}[tbp]
\subfloat[]{\begin{minipage}[c]{0.5\linewidth}
\includegraphics[width=3.2 in, height=3.1 in]{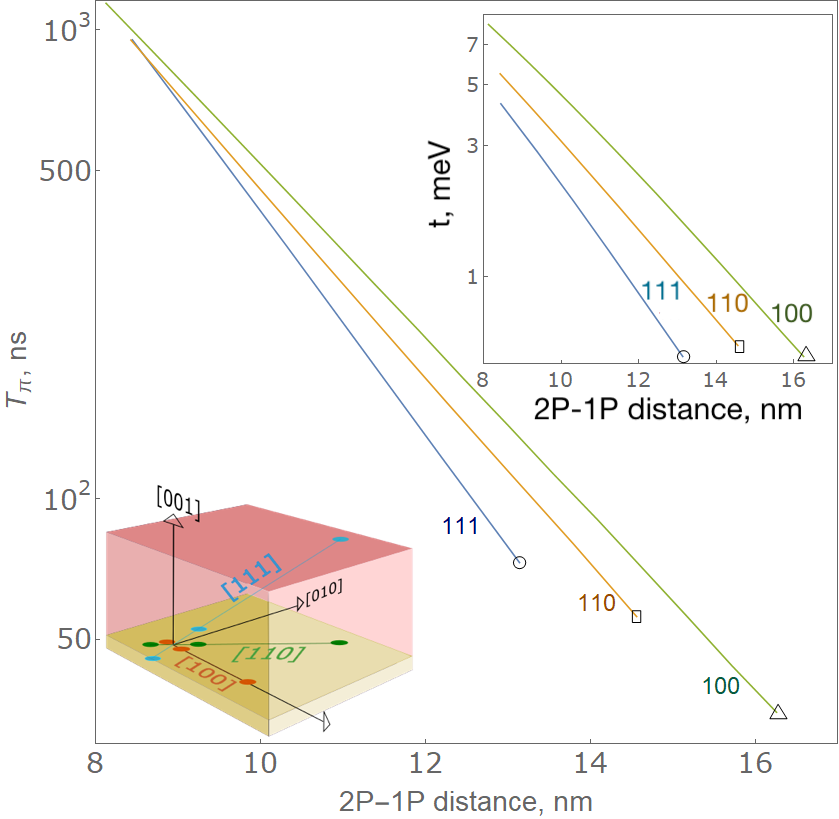}
\label{fig:EDSR1}
\end{minipage}}
\hfill
\subfloat[]{\begin{minipage}[c]{0.5\linewidth}
\includegraphics[width=3.2 in, height=3.1 in]{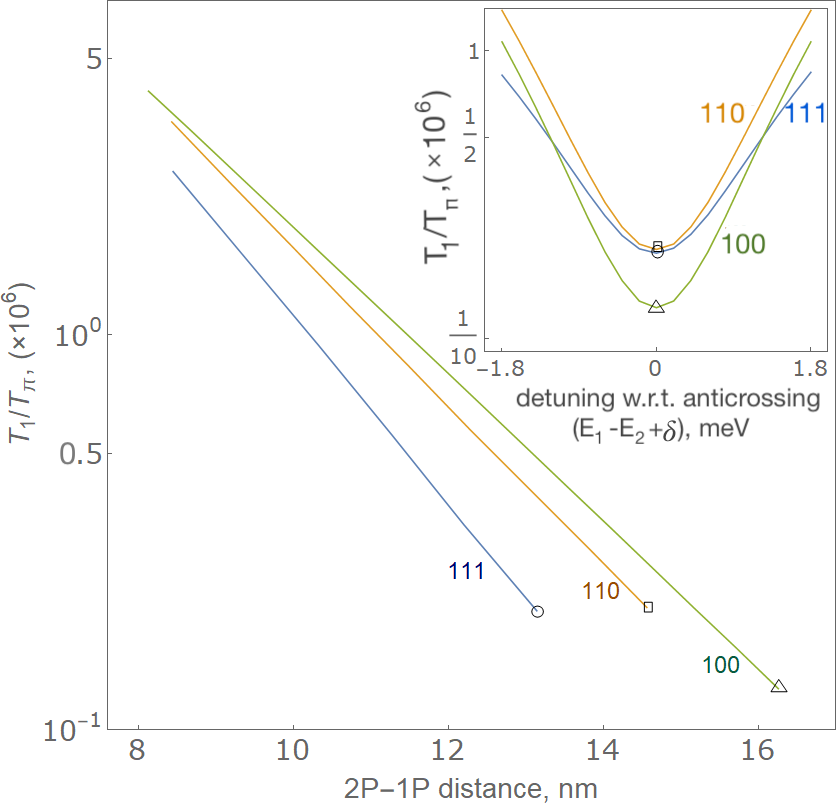}
\label{fig:EDSR2}
\end{minipage}}
\caption{\textbf{2P:1P EDSR time $T_\pi$ and Rabi ratio ($T_1/T_\pi$) vs. 2P-1P separation} at the respective anti-crossings for $[100]\,(-137\,meV)$, $[110]\,(-122.9\,meV)$ and $[111]\,(-111.7\,meV)$ qubit orientations. a) Variation in EDSR spin-flip time with 2P-1P separation. \textbf{inset:} The difference in tunnel coupling for the three 2P-1P orientations are shown (above). The schematic of the three 2P-1P orientations are shown (below). b) Rabi ratio $T_1/T_\pi$ is shown to highlight the number of spin-flip operations possible in a relaxation time. A fixed $B= 1T$ magnetic field is applied perpendicular to the plane of qubit orientation giving a Zeeman splitting of 0.116 meV. In all panels, the position of the triangle ([100]), square ([110]) and circle ([111]) mark the 2P-1P separation where our perturbative approach breaks down. \textbf{inset:} $T_1/T_\pi$ is plotted against the \textit{dc} detuning field $\delta$ in the vicinity of the anticrossing.}
	\label{fig:relax1}
\end{figure*}


\subsection{Hyperfine mediated EDSR and relaxation}

In the vicinity of the charge anti-crossing, the hyperfine interaction enables an electron spin-flip transition $(1,0)\uparrow, \downarrow \leftrightarrow (0,1)\downarrow, \uparrow$ between the left and right dots. If the ac electric field is in resonance with the qubit Zeeman splitting, an electron spin flip takes place owing to the time dependent modulation of the difference between the in-plane hyperfine interactions on the left and right dots, a mechanism analogous to that observed using a micromagnet, \cite{Veldhorst2015two, Kalra2014, Deng2005, assali2011hyperfine, zwanenburg2013silicon} termed hyperfine mediated electron dipole spin resonance (EDSR).\cite{laird2007hyperfine} To describe the effect of a driving AC electric field in the plane of the qubit, the Schrieffer-Wolff (SW) transformation is applied to Eqn.~\ref{eqn:masterhamiltonian} after rotating the qubit basis, which enables us to obtain a low-energy, $2\times 2$ effective qubit Hamiltonian (See supplementary material). The effective Zeeman splitting $\zeta$ due to an applied external magnetic field and out-of-plane hyperfine coupling is much smaller than the orbital energy splitting $\delta\varepsilon\,{=}\,\sqrt{\epsilon^2+4t^2}$, where $\epsilon\,{=}\,e_l-e_r+\delta$ and $t$ denotes $2P-1P$ tunneling. Using $2\zeta\,{\ll}\,\delta\varepsilon$, a condition which boils down to $\zeta\,{\ll}\,t$ at the charge anticrossing, we expand up to second order in small terms. The EDSR rate for a $\pi$-rotation is given by the off-diagonal term $\tilde{H}_{12}$ between $\overline{G}\Uparrow$ and $\overline{G}\Downarrow$ as follows,
	\begin{eqnarray}\label{eqn:EDSR}
	\frac{1}{T_\pi}= \left|\frac{ v_{ac}b\,k_{ge}}{\delta\varepsilon} \right|
	\end{eqnarray}
with $b\,{=}\,\sqrt{b_-b_+}$. The same matrix element $\tilde{H}_{12}$ enables relaxation via coupling to phonons, \cite{borhani2010two, weber2018spin} as described in the supplementary material. The EDSR rate is independent of the qubit Zeeman splitting, a fundamental difference between EDSR mechanisms based on the hyperfine and intrinsic spin-orbit interactions.\cite{rashba2008theory, huang2020impact} For the results in Fig.~\ref{fig:relax1} the $2P:1P$ qubit is operated at the charge anticrossing where the EDSR time is $T_\pi\,{=}\,\frac{2t}{v_{ac}bk_{ge}}$. Following Eqn.~\ref{eqn:EDSR}. $T_\pi\,{\propto}\,t$ implies that at larger 2P-1P separations EDSR is faster, as 2P-1P tunneling decreases.
Nevertheless our SW approximation breaks down beyond a certain value of the 2P:1P separation 
where 2P-1P tunneling is very small and the condition $\zeta\,{\ll}\,t$ is no longer satisfied.

The nuclear spins are polarised in the direction of the external magnetic field i.e. the $z$-direction; and in an ensemble-averaged experiment the in-plane $b_+$, $b_-$ would average to zero leading to a washing out of the Rabi oscillations.\cite{laird2007hyperfine} To have non-zero transverse components one needs to apply a $\frac{\pi}{2}$ pulse to the $1P$ nuclear spin.

We consider three possible orientations of the 2P:1P qubit axis,- along $[110]$, $[110]$ and $[111]$ respectively. We have found that changing the direction of the $2P$ axis alters the results by $1-2\%$, which in practice has no visible effect on the qubit EDSR and relaxation. As the valley compositions of $2P\,{\parallel}\,100$, $2P\,{\parallel}\,110$ and $2P\,{\parallel}\,111$ are not drastically different, for fixed orientation of the 1P donor dot the overall 2P-1P tunneling matrix does not change much. On the other hand, rotating the 2P:1P qubit axis means the overlap between same valleys (e.g. $2P_x$ to $1P_x$, $2P_y$ to $1P_y$ etc.) changes considerably, which produces significant changes to the overall tunneling matrix elements. Therefore, for simplicity, in the three 2P-1P configurations considered here we take the $2P$ axis to be oriented along the same direction as the $2P:1P$ qubit axis (see inset to Fig.~\ref{fig:EDSR1}). Fig.~\ref{fig:relax1} shows the effect of varying the orientation of the $2P:1P$ qubit axis, as well as the effect of changing $2P-1P$ separation on the EDSR time $T_\pi$ and \textit{Rabi ratio} of the relaxation time $T_1$ to the gate time $T_\pi$, all evaluated at the charge anti-crossing. The number of gate operations per relaxation time decreases as the dots are further apart. We consider first having the $2P:1P$ axis $\parallel [100]$. For a constant 2P-1P separation of $13$ nm, the 2P-1P tunnel coupling takes the value $t=1.7$ meV, with valley contributions $t_{\pm x}\,{=}\,0.31$ meV, $t_{\pm y}\,{=}\,0.26$ meV, $t_{\pm z}\,{=}\,0.26 meV$. The magnitude of the $2P -1P$ tunnel coupling is at its highest for 2P:1P$\parallel$[100]. We find that the $x$-valleys make the dominant contribution to $t$ (Fig.~\ref{fig:EDSR1}). Along this direction the wave function overlap between $2P$ and $1P$, and with it the tunnel coupling, is maximal. Since $T_\pi$ and $T_1/T_\pi$ are both $\propto t$, this means the EDSR gate time is longest along this direction, but the Rabi ratio is also at its largest. 
Next, having the $2P:1P$ axis $\parallel[110]$ results in the $x$- and $y$-valleys making a higher contribution to tunneling than the $z$-valleys, namely $t_{\pm x}\,{=}\,0.20$ meV, $t_{\pm y}\,{=}\,0.20$ meV, $t_{\pm z}\,{=}0.11 meV\,$, yet the total value of the tunnel coupling decreases as compared to $[100]$, $t\,{=}\,1$ meV. This is because the overlap of the $2P$ and $1P$ wave functions is smaller than it is along $[100]$. Hence the Rabi ratio $T_1/T_\pi$ as well as the EDSR time $T_\pi$ have smaller values along $[110]$ than along $[100]$. When the qubit axis is $\parallel$[111] all valleys contribute equally: $t_{\pm x}\,{=}\,0.08$ meV, $t_{\pm y}\,{=}\,0.08$ meV, $t_{\pm z}\,{=}0.08 meV\,$, with the total 2P-1P tunneling $t\,{=}\,0.5$ meV, lowest among the three orientations; so the Rabi ratio further decreases (Fig.~\ref{fig:EDSR2}) with EDSR being the fastest for the 2P:1P$\parallel$[111]. Fig.~\ref{fig:EDSR2} also shows that more gate operations within the relaxation time are possible if the qubit is operated away from the anti-crossings of the respective qubit geometries. Due to the difference in 2P-1P coupling for different orientations, the validity of perturbation theory, i.e. $\zeta\,{\ll}\,t$, is satisfied up to different 2P-1P separations. The perturbative analysis is valid up to a 2P-1P separation of $16.3$ nm for [100], $14.6$ nm for [110] and $13.1$ nm for [111]. 

We have thus established that a $2P:1P$ qubit can exhibit fast hyperfine-mediated EDSR and over a million electrical operations within one relaxation time, and that, owing to the difference in interdot tunnelling for the three orientations investigated, the performance is best when the qubit axis is $\parallel$[100].


  \begin{figure*}[t]
\subfloat[]{\begin{minipage}[c]{0.33\linewidth}
\includegraphics[width= \textwidth]{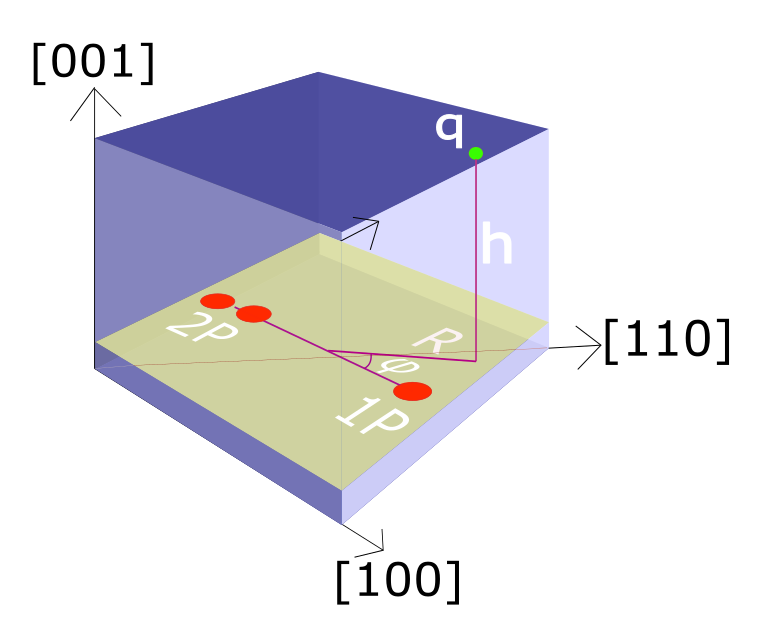}
\label{fig:noise100schematic}
\end{minipage}}
\hfill
\subfloat[]{\begin{minipage}[c]{0.33\linewidth}
\includegraphics[width= \textwidth]{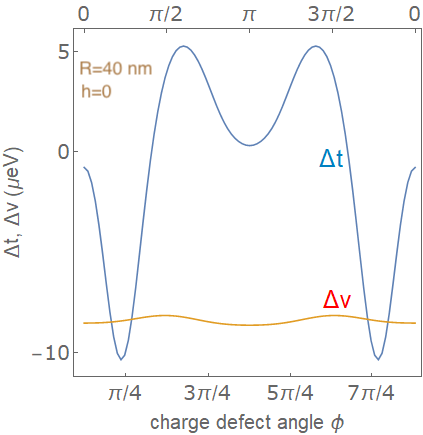}
\label{fig:noise100angle}
\end{minipage}}
\hfill
\subfloat[]{\begin{minipage}[c]{0.33\linewidth}
\includegraphics[width= \textwidth]{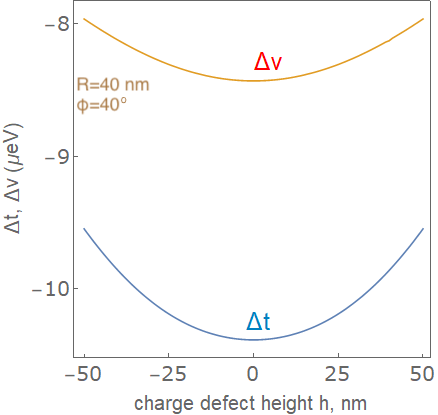}
\label{fig:noise100R}
\end{minipage}}
\hfill
\subfloat[]{\begin{minipage}[c]{0.33\linewidth}
\includegraphics[width=\textwidth]{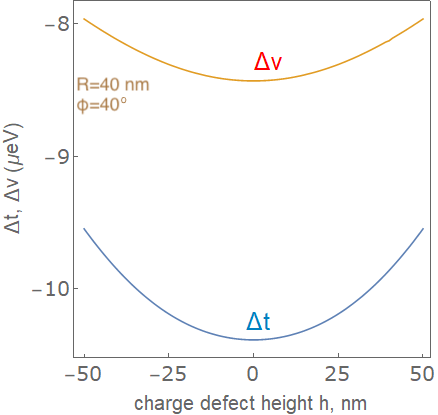}
\label{fig:noise100h}
\end{minipage}}
\hfill
\subfloat[]{\begin{minipage}[c]{0.33\linewidth}
\includegraphics[width=\textwidth]{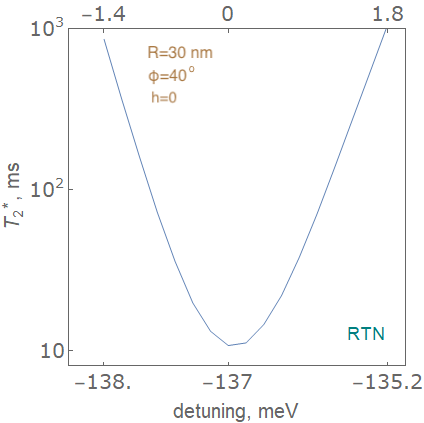}
\label{fig:noise100detRTN}
\end{minipage}}
\hfill
\hfill
\subfloat[]{\begin{minipage}[c]{0.33\linewidth}
\includegraphics[width=\textwidth]{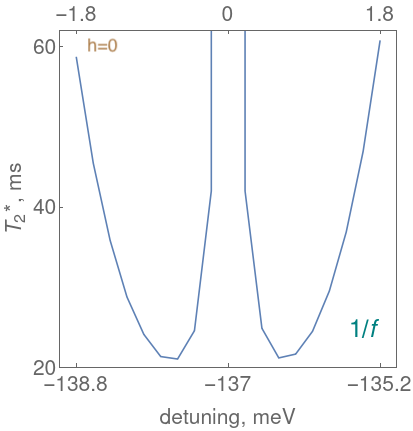}
\label{fig:noise100det1bf}
\end{minipage}}

 \caption{\textbf{$2P:1P\parallel [100]$ qubit fidelity in the presence of charge noise}. a) a schematic of a charge defect $q$ placed at an angle $\phi$ w.r.t. to qubit axis and a distance $R\,$nm away from the center of the qubit i.e. the midpoint between $2P$ and $1P$ dots. The vertical separation of the defect from the qubit plane is $h$. The multi donor quantum dot qubit oriented along $[100]$. For the calculations presented in panel \textit{b-d,f}, the $2P-1P$ distance is $14.4\,$nm; the $2P-1P$ tunneling energy is $t=1\,$meV. b) variation of fluctuations in the tunneling energy (blue) $\Delta t$ and fluctuations in detuning (orange) $\Delta v$ as a function of in-plane ($h=0$) angular orientation $\phi$ of the charge defect, $R=40\,nm$. The oscillation of $\Delta t$ determines in which directions charge noise should be avoided; for all orientations, $\Delta v$ is nearly constant, and comparable to the highest $\Delta t$. c) variation of $\Delta t$ (blue) and $\Delta v$ (orange) as a function of charge defect in-plane distance $R$ from the center of the qubit, $\phi =40^o$. d) variation of $\Delta t$ (blue) and $\Delta v$ (orange) as a function of charge defect height $h$ from the qubit plane; $R=40\,nm,\,\phi =40^o$. e) dephasing time due to random telegraph noise (RTN), $(T^*_{2,RTN})$ as a function of detuning ($\delta$) for random telegraph noise, with an external magnetic field of $1 T$; $R{=}30\,nm,\,\phi{=}40^o,\,h{=}0$. The top label shows the detuning field range to be $\pm2$ meV around the anticrossing. f) the dephasing time ($T^*{_2,1/f}$) due to $1/f$ noise, as a function of detuning ($\delta$). Formally $T^*_{2,1/f}$ tends to infinity here since we have not included higher order terms in the defect potential.} 
 \label{fig:noise100}
 \end{figure*}
 
  \subsection{Nuclear flip-flops and the anisotropic hyperfine interaction}
    
    We now consider the effect of the nuclear field of the three donor nuclear spins (2P+1P) on the electrical operation of the qubit. To begin with, 
    the nuclear dipole-dipole interaction \cite{Deng2005} could induce nuclear-nuclear flip-flops between the three donors independently of the applied electric field. Although the nuclei are relatively close (the two nuclei on the $2P$ dot are $0.5$ nm apart, whereas the nuclei on $1P$ is situated $12$ nm apart from nuclei on $2P$), the time scale for nuclear flip-flops is expected to be 10 ms,\cite{Deng2005} which is very slow compared to the electric field frequency since the rate depends on the product of two nuclear $g$-factors.
    
    We address potential feedback effects on the nucleus in the course of EDSR. One possibility is for the \textit{nuclear} spin to experience electrically driven spin resonance, in which the nuclear spin is flipped by the combination of the ac electric field and inhomogeneous Zeeman field between $2P$-$1P$ due to the electron spin, which enters in the hyperfine interaction. This possibility can be safely discounted, since for it to occur the hyperfine interaction with the electron spin would have to couple the nuclear ground state to orbital excited states of the nucleus. These excited states are extremely high in energy, of the order of MeV.\cite{serge1964nuclei} Moreover, the ac electric field frequency will match the Zeeman splitting of the electron spin states, and will be detuned by three orders of magnitude from the nuclear spin state splitting due to the difference in $g$-factors.

    The electron spin may also have a \textit{back-action} on the nuclear spin, since the electron spin is rotated by EDSR, and this rotation could in principle affect the hyperfine interaction. Flipping the electron spin could result in the hyperfine interaction being also flipped. Nevertheless, as long as the external magnetic field is large enough, this sign flip should have a minimal influence on the nuclear spin. It is worth emphasising that the electron-nuclear system is not allowed to evolve at its natural frequency. Rather, the coefficients $A({\bm R}_i)$ are modified by the electric field fast enough that the electron effectively senses an inhomogeneous magnetic field that depends on its position, analogous to a micromagnet. \cite{han2014graphene}
    
    The anisotropic hyperfine (AHF) interaction between the electron and nuclear spin arising from magnetic dipolar coupling\cite{witzel2007decoherence} can lead to fluctuations in $T_\pi$. Compared to the contact hyperfine Hamiltonian (Eqn.~\ref{eqn:hyperfine}), the AHF Hamiltonian\cite{hale1971calculation} is $H_{AHF}= \overrightarrow{I}\cdot \mathcal{A} \cdot \overrightarrow{S}$, with $\mathcal{A}_{ij}= \frac{8\pi}{3}\gamma_e\gamma_n \hbar^2 \frac{3 x_i x_j- r^2 \delta_{ij}}{r^5}$. The contribution of AHF to the EDSR matrix element is $0.007$ times the contact hyperfine at the anticrossing, \cite{assali2011hyperfine} which will have a negligible effect on qubit operation. 
    
    We conclude that nuclear flip-flops and the anisotropic hyperfine interaction do not play an important part in the electrical operation of $2P:1P$ qubits.


\subsection{Charge noise effects on qubit operation: decoherence and gate errors} 



We assume the sample is isotopically purified, so that there is no random hyperfine field contributing to qubit decoherence. Nevertheless, as compared with single-donor quantum dot qubits, the $2P:1P$ system itself has a dipole moment and is therefore exposed to charge noise. We now discuss qubit decoherence in the presence of this charge noise, which causes certain quantities to fluctuate in the effective $2\times2$ qubit Hamiltonian. \cite{Hung2014} The presence of charge defects in the qubit environment can affect both the $2P$-$1P$ tunnelling energy, producing a fluctuation $\Delta t$ in the tunnelling, as well as in the charge distribution between $2P$ and $1P$ leading to a fluctuation $\Delta v$ in the detuning.\cite{morello2010single, muhonen2015quantifying, chirolli2008decoherence} We focus first on random telegraph noise (RTN), considering a charge defect in the vicinity of the physical qubit, and varying its location.\cite{culcer2013dephasing} The setup is illustrated in Fig.~\ref{fig:noise100schematic}. The defect is characterised by one switching time $\tau$, which we take to be $\tau = 1\mu$s, since experimentally the effect of fluctuators with switching times longer than this can be eliminated by means of dynamical decoupling. For the range of defect distances studied here, the fluctuations $\Delta t$, $\Delta v$ satisfy $\Delta t, \Delta v \ll (\hbar/\tau)$, hence, as described in Ref.~\onlinecite{culcer2013dephasing}, the dephasing rate can be expressed as
	\begin{equation}\label{eqn:RTN}
	\left(\frac{1}{T_2^*}\right)_{RTN} = \frac{32\zeta^2\tau k_{ge}^4}{\hbar^2 \delta\varepsilon^4} \left(\frac{4 t \Delta t}{\delta\varepsilon^2} + \frac{\epsilon \Delta v}{\delta\varepsilon^2}\right)^2,
	\end{equation}
where $\zeta$ is the effective Zeeman splitting, $k_{ge}$ is a matrix element of the hyperfine interaction introduced above, the orbital gap is $\delta\varepsilon$, and $\epsilon= (e_l-e_r+\delta)$. The RTN dephasing time is determined by fluctuations in both the tunnelling and the detuning, and in what follows we study their relative contributions for different defect locations. 
 
Fig.~\ref{fig:noise100angle} shows that $\Delta t$, the fluctuation in the tunnel coupling, is a strong function of the angle $\phi$, which is the azimuthal angle characterising the position vector of the charge defect. It is symmetric about $\pi$, while its absolute value reaches a maximum at $\phi = \pi/4, 7\pi/4$. Interestingly, the integral of $\Delta t$ over the angle $\phi$ is zero, which implies that, if a significant number of defects were randomly located in the vicinity of the qubit, their effect on tunnelling is expected to be washed out. On the other hand, the fluctuation $\Delta v$ in the detuning is nearly constant for all $\phi$. Finally, the variation of $\Delta t$ and $\Delta v$ with charge defect distance $R$ is shown in Fig.~\ref{fig:noise100R}, showing that, as expected, they both decrease with increasing defect-qubit separation at approximately the same rate. The dependence of $\Delta t$ and $\Delta v$ on the vertical position of the charge defect is shown in panel \ref{fig:noise100h}. Whereas there is little overall variation in the range studied, the absolute values of both $\Delta t$ and $\Delta v$ reach their maxima at $h=0,\,R=30\,nm,\,\phi=\pm40^o$, i.e. the worst position for a defect to impact $T_2^*$ is in the same plane as the qubit, at an angle $\pm 40^o$ to the inter-donor axis and at a distance $\simeq\,30\,nm$ away. A charge defect is shown therefore to have a more detrimental effect when placed closer to the $1P$ dot, and, when it lies in the same plane as the qubit. 
 
At the anticrossing, where $\epsilon = 0$, the term $\propto \Delta v$ does not contribute to $T^*_{2,RTN}$, the only contribution stems from the tunneling fluctuation $\Delta t$. Away from the anticrossing the RTN dephasing time increases because the orbital energy gap $\delta\varepsilon$ increases (Fig.~\ref{fig:noise100detRTN}). Both the tunnelling fluctuation $\Delta t $ and detuning fluctuation $\Delta v$ contribute to dephasing, and $T^*_{2,RTN}\propto \delta\varepsilon^8$. 
\begin{table*}[!htbp]
\centering
\begin{tabular}{ l | c | c | c  } 
 \hline
 \hline
  & $T^*_{2,RTN}$ & $T^*_{2,1/f}$& Source\\
 \hline
  single-spin qubit in QD, mediated by Rashba SO  & $3\,ms$ & $20\,\mu s$ & Ref.~\onlinecite{bermeister2014charge}\\
  single-spin qubit in QD, mediated by micromagnet  & $30\,ms$ & $130\,\mu s$  & Ref.~\onlinecite{kha2015micromagnets}\\
 $2P:1P$ qubit at anticrossing $(-137\,meV)$, mediated by hyperfine & $300\,\mu s$ & &Here \\
 $2P:1P$ qubit away from anticrossing $(-135.8\,meV)$, mediated by hyperfine & $55\,ms$ & 30\,ms & Here\\
\hline
\hline
\end{tabular}
\caption{Comparison of dephasing time due to random telegraph noise $T^*_{2,RTN}$ and $1/f$ noise $T^*_{2,1/f}$ for QD single-spin qubits, 2P:1P qubit at the charge anticrossing, and 2P:1P qubit operated away from the anticrossing.}
\label{tab:dephasing}
\end{table*}
The potential of a charge defect makes a substantial contribution to the off-diagonal terms in the qubit Hamiltonian, which are responsible for EDSR. This can lead to EDSR gate errors. To quantify this effect we consider the ratio $\frac{ 1}{T_\pi}\frac{(2\epsilon\Delta v+8t\Delta t)}{\delta\varepsilon^2}$ between the fluctuation-induced term in the Hamiltonian and the EDSR Rabi frequency. At the anticrossing, where only tunneling fluctuations $\Delta t$ contribute, the maximum value of this ratio in the range studied is $0.57$ MHz with a driving electric field of $10\,kV/m$, which corresponds to $2\%$ of the Rabi frequency for our qubit setup. 
This suggests gate errors could be an important effect of random telegraph noise. At the same time, the fractional change in the Rabi frequency could be reduced by simply increasing the driving electric field. 

For $1/f$ noise, which arises from an incoherent superposition of a large number of random telegraph sources, we will assume that the effect of tunneling fluctuations $\Delta t$ is negligible, using the reasoning presented above. On the other hand the detuning is roughly constant as the charge defect location is rotated in the plane so the detuning fluctuations are expected to be important at all times. In light of this, the $1/f$ noise spectrum\cite{shamim2011suppression} $S(\omega)= A k_BT/\omega$, in the qubit subspace, takes the form $S_v(\omega)= \left(\frac{8k_{ge}^2\epsilon \zeta}{\delta\varepsilon^4}\right)^2 S(\omega)$; where $k_{ge}$ is hyperfine energy, $\epsilon=e_l-e_r+\delta$; $e_l,\,e_r$ are the on-site energies of $2P$ and $1P$ dots, $\delta$ is the detuning field, $\zeta$ signify qubit Zeeman splitting and $\delta\varepsilon\,{=}\,\sqrt{\epsilon^2+4t^2}$ is orbital energy splitting respectively. $A\,{=}\,0.1\,\mu eV$ is estimated from experiments. Using $S_x(t)= S_{0x}e^{-\chi(t)}$ (See supplementary material), we obtain:
\begin{equation}\label{eqn:oneoverf}
\left(\frac{1}{T_2^*}\right)_{1/f} \simeq \sqrt{\frac{Ak_BT}{2\hbar^2}}\frac{8 k_{ge}^2\epsilon \zeta}{(\epsilon^2+4t^2)^2}
\end{equation}
 At the anti-crossing the numerator in Eqn.~\ref{eqn:oneoverf} is zero in this approximation as the detuning fluctuation does not contribute, leading to a minimum in dephasing rate. Clearly, when higher-order terms in the defect potential are considered, the value at the anti-crossing will be a small but finite number. Close to the anti-crossing the numerator in Eqn.~\ref{eqn:oneoverf} is dominant resulting in two peaks in the dephasing rate $(T_2^*)^{-1}$ symmetric about the anti-crossing, while further away from the anti-crossing the denominator, which is of higher order in the detuning, dominates the decoherence (Fig.~\ref{fig:noise100det1bf}). In other words, $(T_2^*)_{1/f}$ decreases close to the 2P-1P charge degeneracy point, while increasing further away from anti-crossing. A direct comparison of the coherence properties with single-spin qubits is impossible in the absence of noise data for $2P:1P$ (even for $1P$). Nevertheless, putting together all the above findings suggests the safest operational regime for the $2P:1P$ qubit is away from the anti-crossing, where sensitivity to RTN dephasing and gate errors, as well as to $1/f$ noise, is minimised.
\subsection{Noise sensitivity in $2P:1P$ donor dot qubits vs electrically operated quantum dot qubits}

We discuss briefly the sensitivity to noise in $2P:1P$ donor dot qubits as compared to electrically operated quantum dot qubits. A dipole moment can be induced in quantum dots either by the spin-orbit interaction\cite{bulaev2007electric} or by a nearby micromagnet.\cite{neumann2015simulation,yoneda2014fast}
Dephasing in these architectures was considered in Refs.~\onlinecite{bermeister2014charge} and \onlinecite{kha2015micromagnets} respectively. The sensitivity to noise is determined by two factors: (i) the magnitude of the spin-mixing interaction, which for quantum dots is either the spin-orbit coupling of the magnetic field gradient, while for a $2P:1P$ qubit it is the difference in the hyperfine interaction between the $2P$ and $1P$ sites; and (ii) the asymmetry of the charge distribution, which in a quantum dot is the asymmetry between the ground state and first excited state wave functions, whereas for a $2P:1P$ qubit it is the asymmetry between the $2P$ and $1P$ wave functions.
 
For single-spin qubits in QDs noise introduces a fluctuation between the charge distributions in the ground state and first excited state respectively, leading to a fluctuation $\delta v$ using the notation in Ref.~\onlinecite{bermeister2014charge}. A complete comparison is difficult to make given that dephasing in the quantum dot depends on a number of parameters including the dot radius and aspect ratio. Nevertheless, one comparison can be made straightforwardly: setting the Rashba spin-orbit interaction in a quantum dot qubit equal to the to the hyperfine interaction term $k_{ge}$ in the $2P:1P$ qubit (which is a very realistic assumption), setting the Zeeman splittings to be equal in both qubits, and considering a defect a fixed distance away from either qubit and with the same switching time, we evaluate the resulting $T_2^*$ for RTN and $1/f$ noise. Table.~\ref{tab:dephasing} establishes that noise properties are better for 2P:1P donor configuration than single-spin qubit in QDs, given that the multi-donor dot qubit is operated away from the anticrossing. The qubit orbital energy gap $\delta\varepsilon$ increases in the operational regime away from anticrossing, also the spin-orbit interaction coming from hyperfine decreases, resulting in longer dephasing times.

For $\zeta\,{=}\,60\,\mu eV$, $\tau\,{\approx}\,10^{-6}\,s$, $\delta\varepsilon\,{=}\,1\,meV$, Rashba spin-orbit energy $s_R\,{=}\,1\,\mu eV$, QD radius of $20$ nm, $\delta v\,{=}\,35\,\mu eV$; dephasing time of single-spin qubit is  $T^*_{2,RTN}$=$\,3\,ms.$ With the same parameters and the charge noise sensitivity mediated by hyperfine $k_{ge}$=$\,1\,\mu eV$, fluctuations are higher in a 2P:1P qubit if it is operated at the anticrossing where it is very sensitive to charge noise: $\Delta t\,{=}\,50\,\mu eV$, $\Delta v\,{=}\,10\,\mu eV$ when 2P-1P separation is $16$ nm. Hence the $2P:1P$ dephasing time is shorter ($T^*_{2,RTN}\,{=}\,300\,\mu s$) than that of single-spin qubit. From Eqn.~\ref{eqn:RTN}, $T^*_{2,RTN}\propto \delta\varepsilon^6$ at the anticrossing, similar to a spin qubit subjected to a spin-orbit field by both micromagnet\cite{kha2015micromagnets} and Rashba SOC\cite{bermeister2014charge}in Si QDs.



The orbital splitting dependence for 2P:1P is given by $T^*_{2,1/f}\propto \delta\varepsilon^4$ similar to the micromagnet mediated QD qubit in Ref.~\onlinecite{kha2015micromagnets}. A somewhat different trend is observed for single-spin QD qubits in Ref.~\onlinecite{bermeister2014charge}, where electrical operation relies on the spin-orbit interaction, resulting in $T^*_{2,1/f}\propto \delta\varepsilon^3$. This discrepancy is accounted for by the different energy dependencies between the matrix elements of the spin-orbit interaction and those of the inhomogeneous magnetic field of the micromagnet. 

In summary, $2P:1P$ qubits are more robust against noise than the equivalent electron quantum dot architectures. 

\subsection{Entanglement: comparison of exchange and dipole-dipole coupling} 

\begin{figure}[t]
	\includegraphics[width=3.1 in, height=2.85 in]{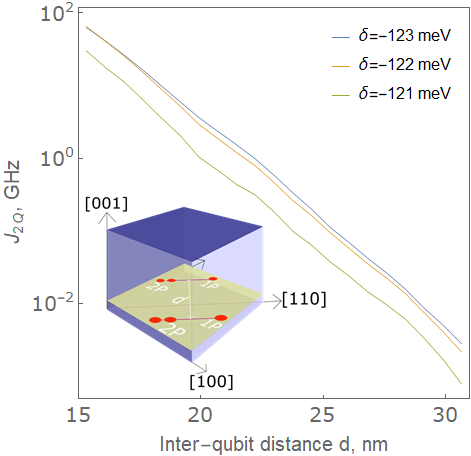}
	\caption{\textbf{Exchange energy ($J_{2Q}$) vs. separation between qubits ($d$) for various bias voltages around the anticrossing value}. The two $2P:1P$ multi donor quantum dot qubits are oriented along $[110]$ and considered for entanglement in \textit{head-to-tail} position at an inter qubit distance $d$ perpendicular to $[110]$(inset).}
	\label{fig:hdtotlex}
\end{figure}

To conclude our discussion, we focus briefly on two possibilities for entangling $2P:1P$ qubits: exchange coupling\cite{loss1998quantum,kane1998silicon}
and dipole-dipole coupling.\cite{trif2007spin,flindt2006spin}
We consider first the exchange energy in the Hund-Mulliken (HM) approximation for two 2P:1P multi-donor quantum dot qubits. Each qubit is oriented along $[110]$, and the qubits are arranged head-to-tail, i.e. perpendicular to $[110]$, as illustrated in Fig.~\ref{fig:hdtotlex}. The ground state for each qubit is denoted by $\overline{G}$ cf. Eqn.~\ref{eqn:masterhamiltonian}. Variation of the exchange energy $J_{2Q}$ with inter-qubit distance $d$ reveals the absence of exchange oscillations when the qubits are operated near the anticrossing (Fig.~\ref{fig:hdtotlex}).
Next we compare the two-qubit exchange coupling with the two-qubit dipole-dipole interaction. Exchange is negligible compared to the direct Coulomb interaction when the distance between the two $2P:1P$ qubits is $\sim 20$ nm. Using the basis of two-particle product states, we can calculate the $16\times 16$ matrix consisting of the direct Coulomb interaction terms \cite{golovach2006electric, salfi2016charge, trif2007spin, tosi2016silicon, flindt2006spin} Then we use the Schrieffer-Wolff (SW) perturbative approach to transform the interaction into the $4\times 4$ two-qubit spin states manifold. Keeping only the dipole terms, the third order SW produces the Ising interaction $H_{SW}^{(3)}= J_{xx}\sigma_{1x}\otimes \sigma_{2x}$; with the coupling strength $J_{xx}=\frac{2k_{ge}^2 Q}{\delta\varepsilon^2}$. The Coulomb matrix elements in the two-qubit basis are given by $\tbkt{mn}{V}{m'n'}$. The diagonal $4\times 4$ sub-matrices of the Coulomb $16\times 16$ matrix are all zero, while the remaining non-zero terms boil down to a single constant $Q$ which is a function of $e\tbkt{\overline{G}}{(\mathbf{r}-\mathbf{R})}{\overline{E}}$, where $e$ is the single electron charge. At $B=1 T$, the gate speed is $0.1$ MHz. Table~ \ref{tab:timescales} compares our $2P:1P$ model to earlier results in terms of operation speed $T_\pi$, relaxation time $T_1$; and also presents our two-qubit gate speeds. In summary, entangling $2P:1P$ qubits using Coulomb exchange is several orders of magnitude faster than using the dipole-dipole interaction. 

\begin{table}
\centering
\begin{tabular}{ l c | c c  } 
 \hline
 \hline
 Parameter & \multicolumn{2}{c}{Value} & Source\\
 \hline
$2P:1P$ EDSR gate time, $T_\pi (ns)$ & $[100]$ & $35$ & \\
 (driving field$\,\approx\,10\,kV/m$)  & $[110]$ & $56$ & Here \\
   & $[111]$ & $73$ & \\
 \hline
 $2P:1P$ EDSR Rabi time, $T_\pi (ns)$ & \multicolumn{2}{c}{$35$} & Ref.~\onlinecite{wang2017all}\\
 (driving field$\,\approx\,15\,kV/m$) &\multicolumn{2}{c}{} &\\
 single donor ESR Rabi time $T_\pi(\mu s)$ & \multicolumn{2}{c}{$20$} & Refs.~\onlinecite{pla2012single,laucht2015electrically} \\
 \hline
 \hline
  & $[100]$ & $0.02$ & \\
 $2P:1P$ relaxation time, $T_1 (s)$ & $[110]$ & $0.04$ & Here \\
 & $[111]$ & $0.05$ & \\
 \hline
 single donor $T_1(s)$ &\multicolumn{2}{c}{$0.07$} & Refs.~\onlinecite{hsueh2014spin,morello2010single,watson2015high}\\
 singlet-triplet qubit $T_1(s)$ &\multicolumn{2}{c}{$0.13$} & Ref.~\onlinecite{gorman2018singlet}\\
\hline
\hline
inter qubit distance, $d (nm)$ & $15$ & $20$ &  \\
\hline
dipole dipole coupling, $J_{xx} (MHz)$ & $0.22$ & $0.09$ & Here \\ 
 exchange energy, $J_{2Q} (GHz)$ & $66$ & $3$ & Here \\ 
 \hline
  $3P-2P$ exchange energy, $J (GHz)$ & \multicolumn{2}{c}{$0.30$} & Ref.~\onlinecite{he2019two} \\ 
 \hline
 \hline
\end{tabular}
\caption{The one qubit time scales: $T_\pi$, $T_1$; and the two qubit energy scales: $J_{2Q}$, $J_{xx}$. In all calculations the external magnetic field B $=1$T.}
\label{tab:timescales}
\end{table}

\section{Summary}
We have developed an analytical wave function for a 2P donor based quantum dot to construct a model of a $2P:1P$ qubit, which is found to be a highly suitable candidate for all-electrical spin quantum computing. Fast EDSR can be achieved in a 2P:1P multi-donor quantum dot with gate times of $10 - 50$ns at electric fields of $10000-50000$ V/m at the charge anti-crossing. The spin relaxation time due to phonons satisifes $1/T_1 \propto B^5$ and, at the anti-crossing, allows in excess of a million qubit operations. Random telegraph noise can lead to sizable fluctuations in tunneling between the $2P$ and $1P$ sites as well as fluctuations in the detuning, which can cause both decoherence and gate errors leading to loss of fidelity. Nevertheless, we have shown that qubits can be immune to RTN and $1/f$ noise some distance away from the charge anti-crossing. Efficient entanglement can be achieved using exchange, which does not exhibit oscillations as a function of qubit separation, while the dipole-dipole interaction is considerably slower. In the future our theoretical method could be extended to $3P$ multi-donor quantum dot configurations and to several qubits in order to examine cross-talk and further issues related to scaling up multi-donor quantum dot qubits. 
\acknowledgments 
 
The authors would like to thank Andre Saraiva for many enlightening discussions. This research is supported by the Australian Research Council Centre of Excellence in Future Low-Energy Electronics Technologies (project number CE170100039), the Australian Research Council Centre of Excellence for Quantum Computation and Communication Technology (project number CE170100012) and Silicon Quantum Computing Pty Ltd. XH is funded by ARO grant no. W911NF1710257. RR acknowledges funding from the U.S. Army Research Oﬃce grant no. W911NF-17-1-0202. MYS acknowledges an Australian Research Council Laureate Fellowship.
 
\bibliography{reference}
\end{document}


\title{Supplementary material: Optimisation of electrically-driven multi-donor quantum dot qubits}
\author{Abhikbrata Sarkar}
	\affiliation{School of Physics, The University of New South Wales, Sydney 2052, Australia}
	\affiliation{ARC Centre of Excellence in Future Low-Energy Electronics Technologies, The University of New South Wales, Sydney 2052, Australia}
	\author{Joel Hochstetter}
	\affiliation{School of Physics, The University of New South Wales, Sydney 2052, Australia}
	\author{Allen Kha}
	\affiliation{School of Physics, The University of New South Wales, Sydney 2052, Australia}
	\author{Xuedong Hu}
	\affiliation{Department of Physics, University at Buffalo, SUNY, Buffalo, NY 14260-1500}
	\author{Michelle Y. Simmons}
	\affiliation{Centre for Quantum Computation and Communication Technology, School of Physics,
		The University of New South Wales, Sydney, New South Wales 2052, Australia}
	\author{Rajib Rahman}
	\affiliation{School of Physics, The University of New South Wales, Sydney 2052, Australia}
	\author{Dimitrie Culcer}
	\affiliation{School of Physics, The University of New South Wales, Sydney 2052, Australia}
	\affiliation{ARC Centre of Excellence in Future Low-Energy Electronics Technologies, The University of New South Wales, Sydney 2052, Australia}
	\date{\today}
	\maketitle
\section{Hund-Mulliken Approach}
Denoting the $1P$ donor dot wavefunction by $\Psi_{1D}(\mathbf{r})$ and the $2P$ multi donor dot ground state wavefunction by $\Psi_{2DGS}(\mathbf{r})$, the Hund-Mulliken wavefunctions for 2P:1P multi donor quantum dot are:
\begin{eqnarray}
    \Psi_{+}(\mathbf{r}) = \left(\Psi_{1D}(\mathbf{r}-\mathbf{X})- g  \Psi_{2DGS}(\mathbf{r}+\mathbf{X})\right)/\sqrt{1-2gS+S^2}\nonumber\\
    \Psi_{-}(\mathbf{r}) = \left(\Psi_{2DGS}(\mathbf{r}+\mathbf{X}) - g  \Psi_{1D}(\mathbf{r}-\mathbf{X}) \right)/\sqrt{1-2gS+S^2}\nonumber\\
\end{eqnarray}
where the 1P donor dot is situated at $\mathbf{X}$ and the 2P multi donor dot is situated at $-\mathbf{X}$. $S$ is the overlap integral and is given by:
\begin{eqnarray}
    S &=&\int d^3r \Psi_{1D}(\mathbf{r}+\mathbf{X}))\Psi_{2DGS}(\mathbf{r}-\mathbf{X})\nonumber\\
    g &=& \frac{1- \sqrt{1- S^2}}{S}\nonumber\\
\end{eqnarray}
The two particle singlet-triplet states are:
\begin{eqnarray}
 \Psi_{s-}(\mathbf{r}_1,\mathbf{r}_2)&=&\Psi_-(\mathbf{r}_1) \Psi_-(\mathbf{r}_2)\nonumber\\
 \Psi_{s+}(\mathbf{r}_1,\mathbf{r}_2)&=&\Psi_+(\mathbf{r}_1) \Psi_+(\mathbf{r}_2)\nonumber\\  
 \Psi_{s0}(\mathbf{r}_1,\mathbf{r}_2)&=&(\Psi_+(\mathbf{r}_1) \Psi_-(\mathbf{r}_2)+\Psi_-(\mathbf{r}_1) \Psi_+(\mathbf{r}_2))/\sqrt{2}\nonumber\\
   \Psi_{t}(\mathbf{r}_1,\mathbf{r}_2)&=&(\Psi_+(\mathbf{r}_1) \Psi_-(\mathbf{r}_2)-\Psi_-(\mathbf{r}_1) \Psi_+(\mathbf{r}_2))/\sqrt{2}\nonumber\\
 \end{eqnarray}
 where $\Psi_{s-}\,\Psi_{s+}\,\Psi_{s0}$ are singlet states, symmetric in $r_1\leftrightarrow r_2$; $\Psi_t$ is the asymmetric triplet state. The orbital Hamiltonian in the basis $\{\Psi_{s-}\,\Psi_{s+}\,\Psi_{s0},\Psi_t\}$
 \begin{equation} 
 H_B=
\begin{pmatrix}
\tbkt{\Psi_{s-}}{h_{tot}}{\Psi_{s-}}& \tbkt{\Psi_{s-}}{h_{tot}}{\Psi_{s+}} &  \tbkt{\Psi_{s-}}{h_{tot}}{\Psi_{s0}} & \tbkt{\Psi_{s-}}{h_{tot}}{\Psi_t}\cr
\tbkt{\Psi_{s+}}{h_{tot}}{\Psi_{s-}} & \tbkt{\Psi_{s+}}{h_{tot}}{\Psi_{s+}}& \tbkt{\Psi_{s+}}{h_{tot}}{\Psi_{s0}} & \tbkt{\Psi_{s+}}{h_{tot}}{\Psi_t} \cr
\tbkt{\Psi_{s0}}{h_{tot}}{\Psi_{s-}} & \tbkt{\Psi_{s0}}{h_{tot}}{\Psi_{s+}} & \tbkt{\Psi_{s0}}{h_{tot}}{\Psi_{s0}} & \tbkt{\Psi_{s0}}{h_{tot}}{\Psi_t}\cr
\tbkt{\Psi_t}{h_{tot}}{\Psi_{s-}} & \tbkt{\Psi_t}{h_{tot}}{\Psi_{s+}} & \tbkt{\Psi_t}{h_{tot}}{\Psi_{s0}} & \tbkt{\Psi_t}{h_{tot}}{\Psi_t} \cr
\end{pmatrix}
\end{equation}
with, $h_{tot}{=}h_l+ h_r + C$,
\begin{eqnarray}
    h_l&=&-\frac{\hbar^2\nabla^2}{2m*}+\sum_{D=2P_L,2P_R}\left(-\frac{e^2}{4\pi\epsilon_0\epsilon_r|\mathbf{r}-\mathbf{R}_D|}\right)\nonumber\\
    h_r&=&-\frac{\hbar^2\nabla^2}{2m*}-\left(\frac{e^2}{4\pi\epsilon_0\epsilon_r|\mathbf{r}-\mathbf{R}_{1P}|}\right)\nonumber\\
    C&=&\frac{e^2}{4\pi\epsilon_0\epsilon_r|\mathbf{r}_1-\mathbf{r}_2|}
\end{eqnarray}
\begin{eqnarray}
&&H_B^{(11)}= \tbkt{\Psi_{s-}}{h_l+h_r+C}{\Psi_{s-}}\nonumber\\
&=& \tbkt{\Psi_-(\mathbf{r}_1) \Psi_-(\mathbf{r}_2)}{h(\mathbf{r}_1)+h(\mathbf{r}_2)+ C(\mathbf{r}_1,\mathbf{r}_2)}{\Psi_-(\mathbf{r}_1) \Psi_-(\mathbf{r}_2)}\nonumber\\
&=& \tbkt{\Psi_-(\mathbf{r}_1)}{h(\mathbf{r}_1)}{\Psi_-(\mathbf{r}_1)}\dbkt{\Psi_-(\mathbf{r}_2)}{\Psi_-(\mathbf{r}_2)}+ \tbkt{\Psi_-(\mathbf{r}_2)}{h(\mathbf{r}_2)}{\Psi_-(\mathbf{r}_2)}\dbkt{\Psi_-(\mathbf{r}_1)}{\Psi_-(\mathbf{r}_1)}\nonumber\\
&&+ \tbkt{\Psi_-(\mathbf{r}_1) \Psi_-(\mathbf{r}_2)}{ \frac{1}{|\mathbf{r}_1-\mathbf{r}_2|}}{\Psi_-(\mathbf{r}_1) \Psi_-(\mathbf{r}_2)}\nonumber\\
&=& 2 e_l + U_1\nonumber\\
\end{eqnarray}
where $e_l$ is the on-site energy of $\Psi_-(\mathbf{r})$. $U_1$ denotes on-site repulsion.
\begin{eqnarray}
&&H_B^{(12)}= \tbkt{\Psi_{s-}}{h_l+h_r+C}{\Psi_{s+}}\nonumber\\
&=& \tbkt{\Psi_-(\mathbf{r}_1) \Psi_-(\mathbf{r}_2)}{h(\mathbf{r}_1)+h(\mathbf{r}_2)+ C(\mathbf{r}_1,\mathbf{r}_2)}{\Psi_+(\mathbf{r}_1) \Psi_+(\mathbf{r}_2)}\nonumber\\
&=&  \tbkt{\Psi_-(\mathbf{r}_1) \Psi_-(\mathbf{r}_2)}{ \frac{1}{|\mathbf{r}_1-\mathbf{r}_2|}}{\Psi_+(\mathbf{r}_1) \Psi_+(\mathbf{r}_2)}\nonumber\\
&=& X\nonumber\\
\end{eqnarray}
\begin{eqnarray}
&&H_B^{(13)} = \tbkt{\Psi_{s-}}{h_l+h_r+C}{\Psi_{s0}}\nonumber\\
&=& \tbkt{\Psi_-(\mathbf{r}_1) \Psi_-(\mathbf{r}_2)}{h(\mathbf{r}_1)+h(\mathbf{r}_2)+ C(\mathbf{r}_1,\mathbf{r}_2)}{(\Psi_+(\mathbf{r}_1) \Psi_-(\mathbf{r}_2)+\Psi_-(\mathbf{r}_1) \Psi_+(\mathbf{r}_2))/\sqrt{2}}\nonumber\\
&=& \frac{1}{\sqrt{2}}\tbkt{\Psi_-(\mathbf{r}_1)}{h(\mathbf{r}_1)}{\Psi_+(\mathbf{r}_1)}\dbkt{\Psi_-(\mathbf{r}_2)}{\Psi_-(\mathbf{r}_2)}+\frac{1}{\sqrt{2}} \tbkt{\Psi_-(\mathbf{r}_2)}{h(\mathbf{r}_2)}{\Psi_+(\mathbf{r}_2)}\dbkt{\Psi_-(\mathbf{r}_1)}{\Psi_-(\mathbf{r}_1)}\nonumber\\
&& + \tbkt{\Psi_-(\mathbf{r}_1) \Psi_-(\mathbf{r}_2)}{ \frac{1}{|\mathbf{r}_1-\mathbf{r}_2|}}{(\Psi_+(\mathbf{r}_1) \Psi_-(\mathbf{r}_2)+\Psi_-(\mathbf{r}_1) \Psi_+(\mathbf{r}_2))/\sqrt{2}}\nonumber\\
&=& \sqrt{2}t_b+ w_1\\
&&H_B^{(14)}=0\\
&&H_B^{(22)}=\tbkt{\Psi_{s+}}{h_l+h_r+C}{\Psi_{s+}}\nonumber\\
&=& \tbkt{\Psi_+(\mathbf{r}_1) \Psi_+(\mathbf{r}_2)}{h(\mathbf{r}_1)+h(\mathbf{r}_2)+ C(\mathbf{r}_1,\mathbf{r}_2)}{\Psi_+(\mathbf{r}_1) \Psi_+(\mathbf{r}_2)}\nonumber\\
&=& \tbkt{\Psi_+(\mathbf{r}_1)}{h(\mathbf{r}_1)}{\Psi_+(\mathbf{r}_1)}\dbkt{\Psi_+(\mathbf{r}_2)}{\Psi_+(\mathbf{r}_2)}+ \tbkt{\Psi_+(\mathbf{r}_2)}{h(\mathbf{r}_2)}{\Psi_+(\mathbf{r}_2)}\dbkt{\Psi_+(\mathbf{r}_1)}{\Psi_+(\mathbf{r}_1)}\nonumber\\
&&+ \tbkt{\Psi_+(\mathbf{r}_1) \Psi_+(\mathbf{r}_2)}{ \frac{1}{|\mathbf{r}_1-\mathbf{r}_2|}}{\Psi_+(\mathbf{r}_1) \Psi_+(\mathbf{r}_2)}\nonumber\\
&=& 2 e_r + U_2\nonumber\\
\end{eqnarray}
$t_b$ gives the tunneling matrix element between $\Psi_+(\mathbf{r})$(left) and $\Psi_-(\mathbf{r})$(right). The on-site energy on the right dot is $e_r$, and the on-site repulsion is $U_2$.
\begin{eqnarray}
&&H_B^{(23)}= \tbkt{\Psi_{s+}}{h_l+h_r+C}{\Psi_{s0}}\nonumber\\
&=& \tbkt{\Psi_+(\mathbf{r}_1) \Psi_+(\mathbf{r}_2)}{h(\mathbf{r}_1)+h(\mathbf{r}_2)+ C(\mathbf{r}_1,\mathbf{r}_2)}{(\Psi_+(\mathbf{r}_1) \Psi_-(\mathbf{r}_2)+\Psi_-(\mathbf{r}_1) \Psi_+(\mathbf{r}_2))/\sqrt{2}}\nonumber\\
&=& \frac{1}{\sqrt{2}}\tbkt{\Psi_+(\mathbf{r}_2)}{h(\mathbf{r}_2)}{\Psi_-(\mathbf{r}_2)}\dbkt{\Psi_+(\mathbf{r}_1)}{\Psi_+(\mathbf{r}_1)}+\frac{1}{\sqrt{2}} \tbkt{\Psi_+(\mathbf{r}_1)}{h(\mathbf{r}_1)}{\Psi_-(\mathbf{r}_1)}\dbkt{\Psi_+(\mathbf{r}_2)}{\Psi_+(\mathbf{r}_2)}\nonumber\\
&& + \tbkt{\Psi_+(\mathbf{r}_1) \Psi_+(\mathbf{r}_2)}{ \frac{1}{|\mathbf{r}_1-\mathbf{r}_2|}}{(\Psi_+(\mathbf{r}_1) \Psi_-(\mathbf{r}_2)+\Psi_-(\mathbf{r}_1) \Psi_+(\mathbf{r}_2))/\sqrt{2}}\nonumber\\
&=& \sqrt{2}t_b+ w_2\nonumber\\
\end{eqnarray}
\begin{eqnarray}
&&H_B^{(24)}=0\\
&&H_B^{(33)}=\tbkt{\Psi_{s0}}{h_l+h_r+C}{\Psi_{s0}}\nonumber\\
&=& \left\langle\frac{\Psi_+(\mathbf{r}_1) \Psi_-(\mathbf{r}_2)+\Psi_-(\mathbf{r}_1) \Psi_+(\mathbf{r}_2)}{\sqrt{2}}\right|(h(\mathbf{r}_1)+h(\mathbf{r}_2)+ C(\mathbf{r}_1,\mathbf{r}_2))\left|\frac{\Psi_+(\mathbf{r}_1) \Psi_-(\mathbf{r}_2)+\Psi_-(\mathbf{r}_1) \Psi_+(\mathbf{r}_2)}{\sqrt{2}}\right\rangle\nonumber\\
&=& e_l+ e_r + \tbkt{(\Psi_+(\mathbf{r}_1) \Psi_-(\mathbf{r}_2)+\Psi_-(\mathbf{r}_1) \Psi_+(\mathbf{r}_2))/\sqrt{2}}{C(\mathbf{r}_1,\mathbf{r}_2)}{(\Psi_+(\mathbf{r}_1) \Psi_-(\mathbf{r}_2)+\Psi_-(\mathbf{r}_1) \Psi_+(\mathbf{r}_2))/\sqrt{2}}\nonumber\\
&=& e_l+e_r+ V_+\\
&&H_B^{(34)}=0\\
&&H_B^{(44)}= \tbkt{\Psi_t}{h_l+h_r+C}{\Psi_t}\nonumber\\
&=& \left\langle\frac{\Psi_+(\mathbf{r}_1) \Psi_-(\mathbf{r}_2)-\Psi_-(\mathbf{r}_1) \Psi_+(\mathbf{r}_2)}{\sqrt{2}}\right|(h(\mathbf{r}_1)+h(\mathbf{r}_2)+ C(\mathbf{r}_1,\mathbf{r}_2))\left|\frac{\Psi_+(\mathbf{r}_1) \Psi_-(\mathbf{r}_2)-\Psi_-(\mathbf{r}_1) \Psi_+(\mathbf{r}_2)}{\sqrt{2}}\right\rangle\nonumber\\
&=& e_l+e_r+ V_-\nonumber\\
\end{eqnarray}
\begin{equation}\label{eqn:orbital4x4}
 H_B= e_l+e_r+
\begin{pmatrix}
U_1-(e_r-e_l)& X &  t_b\sqrt{2}+ w_1 & 0\cr
X & U_2+(e_r-e_l) & t_b\sqrt{2}+ w_2 & 0 \cr
t_b\sqrt{2}+w_1 & t_b\sqrt{2}+w_2 & V_+ & 0\cr
0 & 0 & 0 & V_- \cr
\end{pmatrix}
\end{equation}
The exchange energy is given by the the difference between the two lowest eigenvalues of the 4x4 orbital Hamiltonian in Eqn.~\ref{eqn:orbital4x4}.
\section{Effective EDSR Hamiltonian}
With a dc bias $\delta$ applied to the 2P multi donor dot the single electron orbital Hamiltonian in the basis of $\Psi_{-}\uparrow$, $\Psi_{-}\downarrow$, $\Psi_{+}\uparrow$ and $\Psi_{+}\downarrow$ is:
 \begin{equation}
H_0 = \begin{pmatrix}
e_l+ \delta & 0 & | & t & 0 \cr
0 & e_l+ \delta & | & 0 & t \cr
---&---&-|-&---&---&\cr
t & 0 & | & e_r & 0 \cr
0 & t & | & 0 & e_r \cr
\end{pmatrix}
\end{equation}
The eigenvalues of this $4\times 4$ matrix are,- 
  \begin{eqnarray}
  \frac{1}{2} \left( e_l + e_r + \delta - \sqrt{e_l^2 - 2 e_l e_r + e_r^2 + 4 t^2 + 2 e_l \delta - 2 e_r \delta + \delta^2} \right)\nonumber\\
   \frac{1}{2} \left( e_l + e_r + \delta - \sqrt{e_l^2 - 2 e_l e_r + e_r^2 + 4 t^2 + 2 e_l \delta - 2 e_r \delta + \delta^2} \right)\nonumber\\
   \frac{1}{2} \left( e_l + e_r + \delta + \sqrt{e_l^2 - 2 e_l e_r + e_r^2 + 4 t^2 + 2 e_l \delta - 2 e_r \delta + \delta^2} \right)\nonumber\\
   \frac{1}{2} \left( e_l + e_r + \delta + \sqrt{e_l^2 - 2 e_l e_r + e_r^2 + 4 t^2 + 2 e_l \delta - 2 e_r \delta + \delta^2} \right)\nonumber\\
  \end{eqnarray}
 The new basis states which will diagonalise the Hamiltonian,-
 \begin{eqnarray}
  \overline{E} \uparrow = \left(\frac{e_l- e_r + \delta - \sqrt{e_l^2 - 2 e_l e_r + e_r^2 + 4 t^2 + 2 e_l \delta - 2 e_r \delta + \delta^2}}{\sqrt{\left(e_l- e_r + \delta - \sqrt{e_l^2 - 2 e_l e_r + e_r^2 + 4 t^2 + 2 e_l \delta - 2 e_r \delta + \delta^2}\right)^2+ 4t^2}}  \Psi_{-} \right.\nonumber\\
  + \left.\frac{2t}{\sqrt{\left(e_l- e_r + \delta - \sqrt{e_l^2 - 2 e_l e_r + e_r^2 + 4 t^2 + 2 e_l \delta - 2 e_r \delta + \delta^2}\right)^2+4 t^2}} \Psi_{+} \right) \uparrow\nonumber\\
  \end{eqnarray}
  \begin{eqnarray}
  \overline{E} \downarrow = \left(\frac{e_l- e_r + \delta - \sqrt{e_l^2 - 2 e_l e_r + e_r^2 + 4 t^2 + 2 e_l \delta - 2 e_r \delta + \delta^2}}{\sqrt{\left(e_l- e_r + \delta - \sqrt{e_l^2 - 2 e_l e_r + e_r^2 + 4 t^2 + 2 e_l \delta - 2 e_r \delta + \delta^2}\right)^2+ 4t^2}}  \Psi_{-} \right.\nonumber\\
  + \left.\frac{2t}{\sqrt{\left(e_l- e_r + \delta - \sqrt{e_l^2 - 2 e_l e_r + e_r^2 + 4 t^2 + 2 e_l \delta - 2 e_r \delta + \delta^2}\right)^2+4 t^2}} \Psi_{+} \right) \downarrow\nonumber\\
  \end{eqnarray}
  \begin{eqnarray}
  \overline{G} \uparrow = \left(\frac{e_l- e_r + \delta 
  + \sqrt{e_l^2 - 2 e_l e_r + e_r^2 + 4 t^2 + 2 e_l \delta - 2 e_r \delta + \delta^2}}{\sqrt{\left(e_l- e_r + \delta + \sqrt{e_l^2 - 2 e_l e_r + e_r^2 + 4 t^2 + 2 e_l \delta - 2 e_r \delta + \delta^2}\right)^2+ 4t^2}}  \Psi_{-} \right.\nonumber\\
  + \left.\frac{2t}{\sqrt{\left(e_l- e_r + \delta + \sqrt{e_l^2 - 2 e_l e_r + e_r^2 + 4 t^2 + 2 e_l \delta - 2 e_r \delta + \delta^2}\right)^2+4 t^2}} \Psi_{+} \right) \uparrow\nonumber\\
  \end{eqnarray}
  \begin{eqnarray}
  \overline{G} \downarrow = \left(\frac{e_l- e_r + \delta + \sqrt{e_l^2 - 2 e_l e_r + e_r^2 + 4 t^2 + 2 e_l \delta - 2 e_r \delta + \delta^2}}{\sqrt{\left(e_l- e_r + \delta + \sqrt{e_l^2 - 2 e_l e_r + e_r^2 + 4 t^2 + 2 e_l \delta - 2 e_r \delta + \delta^2}\right)^2+ 4t^2}}  \Psi_{-} \right.\nonumber\\
  + \left.\frac{2t}{\sqrt{\left(e_l- e_r + \delta + \sqrt{e_l^2 - 2 e_l e_r + e_r^2 + 4 t^2 + 2 e_l \delta - 2 e_r \delta + \delta^2}\right)^2+4 t^2}} \Psi_{+} \right) \downarrow\nonumber\\
 \end{eqnarray} 
The total hyperfine mediated EDSR Hamiltonian $H_q$ in the $2P:1P$ orbital+spin basis $\{\overline{G}\uparrow$, $\overline{G}\downarrow$,$ \overline{E}\uparrow$, $\overline{E}\downarrow\}$ reads:
\begin{equation}
H_q = \begin{pmatrix}
e_1+\epsilon_z&k_{gg}b_- &|&  k_{ge}b_z+ v_{ac} & k_{ge}b_- \cr
k_{gg}b_+ &e_1-\epsilon_z &|& k_{ge}b_+ & v_{ac}-k_{ge}b_z \cr
-----&-----&|&-----&-----&\cr
k_{eg}b_z+v_{ac} & k_{eg}b_- &|& e_2+\epsilon_z & k_{ee}b_- \cr
k_{eg}b_+ & v_{ac}-k_{eg}b_z &|&k_{ee}b_+ &e_2-\epsilon_z \cr
\end{pmatrix}
\end{equation}
The matrix elements $k_{gg}\,{=}\,\mathcal{A}_0\tbkt{\overline{G}}{\delta(\mathbf{r})}{\overline{G}}$, $k_{ge}\,{=}\,k_{eg}\,{=}\,\mathcal{A}_0\tbkt{\overline{G}}{\delta(\mathbf{r})}{\overline{E}}$ arise from the contact hyperfine interaction; $b_+\,{=}\,\langle I_+ \rangle,\,b_-\,{=}\,\langle I_-\rangle,\,b_z\,{=}\,\langle I_z\rangle$ are the expectation values of the nuclear spin angular momentum operators. $k_{gg}b_z$ stems from the out-of-plane hyperfine interaction, and this is simply added to the total Zeeman splitting $e_z$, modifying the spin states splitting as $\epsilon_z\,{=}\,e_z+k_{gg}b_z$. First, the upper diagonal $2\times 2$ block of the $H_{hf}$ is considered. 
 \begin{equation}
 H_{hf}^{2\times 2} = \begin{pmatrix}
e_1+\epsilon_z&k_{gg}(b_x-i b_y) \cr
k_{gg}(b_x^*+i b_y^*) &e_1-\epsilon_z  \cr
\end{pmatrix}
\end{equation} 
Diagonalisation of $H_{hf}^{2\times 2}$ gives us a new set of basis states with spin up and down combination, i.e. a \textit{rotated} basis:

 \begin{eqnarray}
  \tilde{a}= \alpha \overline{G}\uparrow + \gamma \overline{G}\downarrow \nonumber , \tilde{b}=\beta \overline{G}\uparrow+ \delta \overline{G}\downarrow \nonumber ,
  \tilde{c}=\alpha \overline{E}\uparrow +\gamma \overline{E}\downarrow \nonumber ,
  \tilde{d}=\beta \overline{E}\uparrow+\delta \overline{E}\downarrow
\end{eqnarray}
where, 
\begin{eqnarray}\label{eq:coeffsbasis}
 \alpha= \frac{\epsilon_z- \sqrt{\epsilon_z^2+(b_x^2+b_y^2)k_{gg}^2}}{\sqrt{(b_x^2 + 
 b_y^2)k_{gg}^2 +\left(\epsilon_z- \sqrt{\epsilon_z^2+(b_x^2+b_y^2)k_{gg}^2}\right)^2}}\nonumber ,\beta= \frac{\epsilon_z+\sqrt{\epsilon_z^2+(b_x^2+b_y^2)k_{gg}^2}}{\sqrt{(b_x^2 + 
 b_y^2)k_{gg}^2 +\left(\epsilon_z+ \sqrt{\epsilon_z^2+(b_x^2+b_y^2)k_{gg}^2}\right)^2}}\nonumber\\
\gamma= \frac{k_{gg} (b_x+ i b_y)}{\sqrt{(b_x^2 + 
 b_y^2)k_{gg}^2 +\left(\epsilon_z- \sqrt{\epsilon_z^2+(b_x^2+b_y^2)k_{gg}^2}\right)^2}}\nonumber ,
\delta= \frac{k_{gg} (b_x+ i b_y)}{\sqrt{(b_x^2 + 
 b_y^2)k_{gg}^2 +\left(\epsilon_z+ \sqrt{\epsilon_z^2+(b_x^2+b_y^2)k_{gg}^2}\right)^2}}\nonumber\\
\end{eqnarray}
 The hyperfine strength $\mathcal{A}_0$ depends on the detuning $\delta$, which can be checked by varying $\delta$ in the qubit ground state $\overline{G}$ and excited state $\overline{E}$ wavefunctions. Near the anticrossing, $\mathcal{A}_0$ has a linear dependence on detuning and allows the charge transition $(1,0) \leftrightarrow (0,1)$ between left and right dot within $10 kV/m$ electric field range.
  An electric dipole transition can be driven by an ac electric field $\widetilde{E}(t)$ applied between $G1$ and $G2$, modifying $\mathcal{A}_0$ with time. If the ac electric field is in resonance with the qubit Zeeman splitting, a spin flip takes place due to the difference between the hyperfine energy of the left and right dot, a mechanism that has also been observed using a micromagnet. This is known as the hyperfine mediated electron dipole spin resonance (EDSR). Turning on the ac field, the EDSR Hamiltonian in the rotated basis is given by:
\begin{equation}\label{eq:transfEDSR}
H'= \begin{pmatrix}
e_1+\zeta & 0 & | & P_1 & Q_1 \cr
0 & e_1-\zeta & | & Q_2 & P_2 \cr
---&---&-|-&---&---&\cr
P_1^\dag &Q_2^\dag & | & e_2+\zeta & 0 \cr
Q_1^\dag & P_2^\dag & | & 0 & e_2-\zeta\cr
\end{pmatrix}.
\end{equation}
where, $\zeta=\sqrt{\epsilon_z^2+(b_x^2+b_y^2)k_{gg}^2}$. The matrix elements are calculated below:
\begin{eqnarray}
H'_{13}= P_1 &=& \bra{\tilde{a}^\dag}H'\ket{\tilde{c}}\nonumber\\
&=& \bra{\alpha \overline{G}\uparrow+\gamma^\dag \overline{G}\downarrow}H'\ket{\alpha \overline{E}\uparrow+\gamma \overline{E}\downarrow}=\bra{\alpha \overline{G}\uparrow+\gamma^\dag \overline{G}\downarrow}(H_{el}+H_{hf})\ket{\alpha \overline{E}\uparrow+\gamma \overline{E}\downarrow}\nonumber\\
&=&\bra{\overline{G}}\bra{\alpha\uparrow+\gamma^\dag\downarrow}(H_{el}+H_{hf})\ket{\alpha\uparrow+\gamma\downarrow}\ket{\overline{E}}\nonumber\\
&=&\bra{\overline{G}}\bra{\alpha\uparrow+\gamma^\dag\downarrow}H_{el}\ket{\alpha\uparrow+\gamma\downarrow}\ket{\overline{E}}+\bra{\overline{G}}\bra{\alpha\uparrow+\gamma^\dag\downarrow}H_{hf}\ket{\alpha\uparrow+\gamma\downarrow}\ket{\overline{E}}\nonumber\\
&=&\bra{\overline{G}}e E_0 x\ket{\overline{E}}\dbkt{\alpha\uparrow+\gamma^\dag\downarrow}{\alpha\uparrow+\gamma\downarrow}\nonumber\\
&&+\bra{\overline{G}}\delta(r)\ket{\overline{E}}\bra{\alpha\uparrow+\gamma^\dag\downarrow}( b_x\sigma_x + b_y \sigma_y +b_z\sigma_z)\ket{\alpha\uparrow+\gamma\downarrow}\nonumber\\
&=&v_{ac}\dbkt{\alpha\uparrow+\gamma^\dag\downarrow}{\alpha\uparrow+\gamma\downarrow}+k_{ge}\bra{\alpha\uparrow+\gamma^\dag\downarrow}( b_x\sigma_x + b_y \sigma_y +b_z\sigma_z)\ket{\alpha\uparrow+\gamma\downarrow}\nonumber\\
&=&v_{ac}(\alpha^2+\gamma^2)+k_{ge}\left( b_x (\alpha\gamma+\alpha\gamma^\dag) + i b_y(\alpha\gamma^\dag-\alpha\gamma) +b_z(\alpha^2-\gamma^2)\right)\nonumber\\
&=&
v_{ac}+ (\alpha^2 - \gamma\gamma^\dag)k_{ge}b_z +\alpha\gamma^\dag k_{ge} (b_x + i b_y)
 +\alpha \gamma k_{ge}(b_x - i b_y)
\end{eqnarray}
$(\alpha^2+\gamma^2)=1$ can be deduced straightforward. 
\begin{eqnarray}
 \alpha^2-|\gamma|^2 &=& \left(\frac{\epsilon_z- \sqrt{\epsilon_z^2+(b_x^2+b_y^2)k_{gg}^2}}{\sqrt{(b_x^2 + 
 b_y^2)k_{gg}^2 +\left(\epsilon_z- \sqrt{\epsilon_z^2+(b_x^2+b_y^2)k_{gg}^2}\right)^2}}\right)^2\nonumber\\
 &&-\left |\frac{k_{gg} (b_x- i b_y)}{\sqrt{(b_x^2 + 
 b_y^2)k_{gg}^2 +\left(\epsilon_z- \sqrt{\epsilon_z^2+(b_x^2+b_y^2)k_{gg}^2}\right)^2}}\right |^2\nonumber\\
 &=& \frac{\left(\epsilon_z- \sqrt{\epsilon_z^2+(b_x^2+b_y^2)k_{gg}^2}\right)^2-(b_x^2+b_y^2)k_{gg}^2}{(b_x^2 + 
 b_y^2)k_{gg}^2 +\left(\epsilon_z- \sqrt{\epsilon_z^2+(b_x^2+b_y^2)k_{gg}^2}\right)^2}\nonumber\\
  &=& \frac{\epsilon_z^2 - 2\epsilon_z \sqrt{\epsilon_z^2+(b_x^2+b_y^2)k_{gg}^2}+\epsilon_z^2+(b_x^2+b_y^2)k_{gg}^2-(b_x^2+b_y^2)k_{gg}^2}{(b_x^2 + 
 b_y^2)k_{gg}^2 +\epsilon_z^2 - 2\epsilon_z \sqrt{\epsilon_z^2+(b_x^2+b_y^2)k_{gg}^2}+\epsilon_z^2+(b_x^2+b_y^2)k_{gg}^2}\nonumber
 \end{eqnarray}
 \begin{eqnarray}
  &=& \frac{2\epsilon_z^2 - 2\epsilon_z \sqrt{\epsilon_z^2+(b_x^2+b_y^2)k_{gg}^2}}{2(b_x^2 + 
 b_y^2)k_{gg}^2 +2\epsilon_z^2 - 2\epsilon_z \sqrt{\epsilon_z^2+(b_x^2+b_y^2)k_{gg}^2}}\nonumber\\
 &=&\frac{-2\epsilon_z \cancel{\left(\sqrt{\epsilon_z^2+(b_x^2+b_y^2)k_{gg}^2}-\epsilon_z\right)}}{2\cancel{\left(\sqrt{\epsilon_z^2+(b_x^2+b_y^2)k_{gg}^2}-\epsilon_z\right)}\sqrt{\epsilon_z^2+(b_x^2+b_y^2)k_{gg}^2}}=\frac{-\epsilon_z}{\zeta} \nonumber\\
 \nonumber
 \end{eqnarray}
 \begin{eqnarray}
 \Rightarrow P_1&=&v_{ac}- \frac{\epsilon_z k_{ge}b_z}{\zeta} +\alpha\gamma^\dag k_{ge} (b_x + i b_y)+\alpha \gamma k_{ge}(b_x - i b_y)\nonumber\\
 &=& v_{ac}- \frac{\epsilon_z k_{ge}b_z}{\zeta}-\frac{\cancel{(\zeta-\epsilon_z)}}{2\cancel{\left(\zeta-\epsilon_z\right)}\zeta} (2(b_x^2+b_y^2))k_{ge}k_{gg}\nonumber\\
 &=& v_{ac}- \frac{\epsilon_z k_{ge}b_z}{\zeta} -\frac{b_x^2+b_y^2}{\zeta}k_{ge}k_{gg}
 \end{eqnarray}
 Similarly,
 \begin{eqnarray}
 H'_{24}=P_2=v_{ac}+ \frac{\epsilon_z k_{ge}b_z}{\zeta} +\frac{b_x^2+b_y^2}{\zeta}k_{ge}k_{gg}
 \end{eqnarray}
 \begin{eqnarray}
H'_{14}= Q_1&=& \bra{\tilde{a}^\dag}H'\ket{\tilde{d}}= \bra{\alpha \overline{G}\uparrow+\gamma^\dag \overline{G}\downarrow}H'\ket{\beta \overline{E}\uparrow+\delta \overline{E}\downarrow}\nonumber\\
&=&\bra{\alpha \overline{G}\uparrow+\gamma^\dag \overline{G}\downarrow}(H_{el}+H_{hf})\ket{\beta \overline{E}\uparrow+\delta\overline{E}\downarrow}\nonumber\\
&=&\bra{\overline{G}}\bra{\alpha\uparrow+\gamma^\dag\downarrow}(H_{el}+H_{hf})\ket{\beta\uparrow+\delta\downarrow}\ket{\overline{E}}\nonumber\\
&=&\bra{\overline{G}}\bra{\alpha\uparrow+\gamma^\dag\downarrow}H_{el}\ket{\beta\uparrow+\delta\downarrow}\ket{\overline{E}}+\bra{\overline{G}}\bra{\alpha\uparrow+\gamma^\dag\downarrow}H_{hf}\ket{\beta\uparrow+\delta\downarrow}\ket{\overline{E}}\nonumber\\
&=&\bra{\overline{G}}e E_0 x\ket{\overline{E}}\dbkt{\alpha\uparrow+\gamma^\dag\downarrow}{\beta\uparrow+\delta\downarrow}\nonumber\\
&&+\bra{\overline{G}}\delta(r)\ket{\overline{E}}\bra{\alpha\uparrow+\gamma^\dag\downarrow}( b_x\sigma_x + b_y \sigma_y +b_z\sigma_z)\ket{\beta\uparrow+\delta\downarrow}\nonumber\\
&=&v_{ac}\dbkt{\alpha\uparrow+\gamma^\dag\downarrow}{\beta\uparrow+\delta\downarrow}+k_{ge}\bra{\alpha\uparrow+\gamma^\dag\downarrow}( b_x\sigma_x + b_y \sigma_y +b_z\sigma_z)\ket{\beta\uparrow+\delta\downarrow}\nonumber\\
&=&k_{ge}\left( b_x (\alpha\delta+\beta\gamma^\dag) + i b_y(\beta\gamma^\dag-\alpha\delta) +b_z(\alpha\beta-\gamma^\dag \delta)\right)\nonumber\\
&=&
(\alpha\beta-\gamma^\dag \delta)k_{ge}b_z +\alpha\delta k_{ge} (b_x - i b_y)
 +\beta \gamma^\dag k_{ge}(b_x + i b_y)\nonumber\\
 &=&\frac{k_{gg}\sqrt{b_x^2+b_y^2}k_{ge} b_z}{\zeta}+ \frac{(\epsilon_z - \zeta)(b_x^2+b_y^2)+(\epsilon_z + \zeta)(b_x^2+b_y^2)}{2\zeta\sqrt{b_x^2+b_y^2}}k_{ge}\nonumber\\
 &=&\frac{k_{gg}\sqrt{b_x^2+b_y^2}k_{ge} b_z}{\zeta}+ \frac{\epsilon_z}{\zeta}\sqrt{b_x^2+b_y^2}k_{ge}\nonumber
\end{eqnarray}
Evidently, $H'_{23}= Q_2 = Q_1 $. $k_{ge} b_z/\zeta \simeq \frac{1}{400}$, so the terms $P_1$, $P_2$, $Q_1$, $Q_2$, to very good approximation, effectively become:
\begin{eqnarray}
P_1 = P_2 \simeq v_{ac},\ \  Q_1 = Q_2 \simeq \sqrt{b_x^2+b_y^2}k_{ge}\nonumber
\end{eqnarray}
The off-diagonal spin-flip term after applying the Schrieffer-Wolff transformation, under the assumption that the \textit{effective} qubit Zeeman splitting is small compared to the orbital energy gap, $\zeta \ll t$:
\begin{eqnarray}\label{eqn:rabifreq}
  \widetilde{H}_{12} &=& \frac{1}{2} H'_{13} H'_{32}\left[\frac{1}{\xi_1-\xi_3}+\frac{1}{\xi_2-\xi_3}\right]+\frac{1}{2} H'_{14} H'_{42}\left[\frac{1}{\xi_1-\xi_4}+\frac{1}{\xi_2-\xi_4}\right]\nonumber\\
   &=& \frac{1}{2}P_1 Q_2^\dag\left[\frac{1}{e_1-e_2}+\frac{1}{e_1-e_2- 2\zeta}\right] + \frac{1}{2} Q_1 P_2^\dag\left[\frac{1}{e_1-e_2}+\frac{1}{e_1-e_2+ 2\zeta}\right]\nonumber\\
  \Omega &=& \frac{v_{ac} |b| k_{ge}}{(e_1-e_2)}=\frac{1}{T_\pi}
\end{eqnarray}
The diagonal elements from Schrieffer-Wolff transformation are given by,
 \begin{eqnarray}
\widetilde{H}_{11} &=& \frac{1}{2} H'_{13} H'_{31}\left[\frac{2}{\xi_1-\xi_3}\right] 
+\frac{1}{2} H'_{14} H'_{41}\left[\frac{2}{\xi_1-\xi_4}\right]\nonumber\\
&=& P_1^2\left[\frac{1}{e_1-e_2}\right]+Q_1^2\left[\frac{1}{e_1-e_2+ 2\zeta}\right]\nonumber\\
\widetilde{H}_{22} &=& \frac{1}{2} H'_{23} H'_{32}\left[\frac{2}{\xi_2-\xi_3}\right] 
+\frac{1}{2} H'_{24} H'_{42}\left[\frac{2}{\xi_2-\xi_4}\right]\nonumber\\
&=& P_2^2\left[\frac{1}{e_1-e_2}\right]+Q_1^2\left[\frac{1}{e_1-e_2- 2\zeta}\right]\nonumber\\
\end{eqnarray} 
The 2x2 effective qubit Hamiltonian is given by $\widetilde{H}_{ij}\,;\,i,j\in\{1,2\}$.

\section{Relaxation time}
  The relaxation time $T_1$ carries information about the coupling of the electron spin with the crystalline and electromagnetic environment. The anisotropy of the deformation potential is taken into account, and the Hamiltonian takes the form:
 \begin{eqnarray}   
      	 H_{e-ph} &=&\sum_\lambda \sum_q -iq (a^\dag_{qs}exp(iq \cdot r)+a_{qs}exp(-iq \cdot r)) \nonumber\\
      	 &&\times\left(\Xi_d \mathbf{e_{\lambda,x}}\hat{q}_x+\Xi_d \mathbf{e_{\lambda,y}}\hat{q}_y+ (\Xi_d+\Xi_u)\mathbf{e_{\lambda,z}}\hat{q}_z \right)\nonumber\\
 \end{eqnarray}
 where $a^\dag_{qs}$ and $a_{qs}$ are the phonon creation and annihilation operators, $q$ is the phonon wavevector, $\hat{\mathbf{e}}_\lambda$ is the phonon polarization vector, the deformation potential is given by the tensor $\Xi_{ij}$:
 \begin{equation}
     \Xi_{ij}=\begin{pmatrix}
     \Xi_d & 0 & 0 \\
     0 & \Xi_d & 0\\
     0 & 0 & \Xi_d+\Xi_u\\
     \end{pmatrix}
 \end{equation}
The initial and final phonon number states produces $\tbkt{n_{q+1}}{a^\dag_{qs}}{n_q} \propto \sqrt{n_q+1}$; in the low temperature region $\sqrt{n_q+1}\sim 1$, as $n_q$ follows Boltzmann distribution. The electron-phonon Hamiltonian can be written as
\begin{eqnarray}
 H_{e-ph} &=& \sum_q \sum_\lambda \frac{-iq}{2} \sqrt{\frac{\hbar}{2NV\rho \omega_{q\lambda}}}exp(iq \cdot r)\times \nonumber\\
 &&\left(\Xi_d \mathbf{e_{\lambda,x}}\hat{q}_x+\Xi_d \mathbf{e_{\lambda,y}}\hat{q}_y+ (\Xi_d+\Xi_u)\mathbf{e_{\lambda,z}}\hat{q}_z \right)\nonumber\\
\end{eqnarray}
where $NV$ denotes the volume of the crystal, $\rho$ is the mass density and $\omega_{q\lambda}$ is the acoustic phonon frequency.
 Using $ \hbar \omega_{q\lambda} = \hbar v_\lambda q $, we obtain:
 \begin{equation}
 \frac{1}{T_1}= \frac{2 \pi}{\hbar}\sum_q\sum_\lambda \left| \tbkt{\overline{G}'\uparrow}{H_{hf}}{\overline{G}'\downarrow} \right|^2 \delta(2\zeta- \hbar v_\lambda q)
 \end{equation}

 The perturbation in our basis states due to presence of vibration would modify the qubit ground state
 \begin{equation}
  \ket{\overline{G}'}= \ket{\overline{G}}+ \frac{\tbkt{\overline{E}}{H_{e-ph}}{\overline{G}}\ket{\overline{E}}}{e_2- e_1}\nonumber
 \end{equation}
 the qubit spin states couling term becomes:
 \begin{equation}
  \tbkt{\overline{G}'\uparrow}{H_{hf}}{\overline{G}'\downarrow}= 2 \frac{\tbkt{\overline{E}}{H_{e-ph}}{\overline{G}}^*}{e_2- e_1} \tbkt{\overline{E}\uparrow}{H_{hf}}{\overline{G}\downarrow}
 \end{equation}

 \begin{eqnarray}
 \frac{1}{T_1}&=& \sum_\lambda\frac{2 \pi}{\hbar} \frac{1}{\hbar v_\lambda}\frac{L^3}{8\pi^3}\frac{\hbar}{2NV\rho v_\lambda} \int_\Omega\ d\Omega_\lambda\int q^3\ dq\ \left|\frac{ \tbkt{\overline{E}}{e^{i \mathbf{q}\cdot\mathbf{r}}}{\overline{G}} \tbkt{\overline{E}\uparrow}{H_{hf}}{\overline{G}\downarrow}}{e_2-e_1}\right|^2 \delta\left(\frac{2\zeta}{\hbar v_\lambda}- q\right)  \nonumber\\
 \end{eqnarray}  
 where $\Omega$ denotes the angular integral. $\tbkt{\overline{G}}{i q\cdot r}{\overline{E}}\simeq \tbkt{\overline{G}}{1+iq\cdot r}{\overline{E}}=\cancelto{0}{\dbkt{\overline{G}}{\overline{E}}}+\tbkt{\overline{G}}{iq\cdot r}{\overline{E}}$; using $\tbkt{\overline{E}\uparrow}{H_{hf}}{\overline{G}\downarrow}\equiv |b| k_{ge}$, and assuming $|b|=1$:
 \begin{eqnarray}
 \frac{1}{T_1}&=& \sum_\lambda \frac{k_{ge}^2}{32\pi^2t^2\hbar v_\lambda^2\rho } \int_\Omega\ d\Omega_\lambda\int q^3\ dq\ q^2\left| \tbkt{\overline{E}}{i \hat{q}\cdot r}{\overline{G}}\right|^2 \delta\left(\frac{2\zeta}{\hbar v_\lambda}- q\right)  \nonumber\\
 &=& \sum_\lambda \frac{k_{ge}^2}{32\pi^2t^2\hbar v_\lambda^2\rho }\int q^5 \delta\left(\frac{2\zeta}{\hbar v_\lambda}- q\right)\ dq \int_\Omega\ d\Omega_\lambda\left| \tbkt{\overline{E}}{\hat{q}\cdot r}{\overline{G}}\right|^2   \nonumber\\
  &=& \sum_\lambda \frac{\zeta^5 k_{ge}^2}{\pi^2t^2\hbar^6 v_\lambda^7\rho }\int_\Omega\ d\Omega_\lambda\left| \tbkt{\overline{E}}{ \hat{q}\cdot r}{\overline{G}}\right|^2   \nonumber\\
 \end{eqnarray}
 final form of relaxation time formula with the angular integral reads:
\begin{equation}
\frac{1}{T_1}=\sum_\lambda \frac{\zeta^5 k_{ge}^2}{\pi^2t^2\hbar^6 v_\lambda^7\rho }\int_0^{2\pi}d\phi\ \int_{-1}^1 d(cos\theta)\left| \tbkt{\overline{E}}{\hat{q}\cdot r}{\overline{G}} \left(\Xi_d \mathbf{e_{\lambda,x}}\hat{q}_x+\Xi_d \mathbf{e_{\lambda,y}}\hat{q}_y+ (\Xi_d+\Xi_u)\mathbf{e_{\lambda,z}}\hat{q}_z\right)\right|^2 
\end{equation}

\textbf{longitudinal polarization}: $v_l = 8480\ m\,s^{-1}$, $\mathbf{\hat{q}}\ ||\ \mathbf{e_l}$, and the components are $\{sin\theta\,cos\phi$, $sin\theta\,sin\phi$, $cos\theta\}$. 

\textbf{transverse polarization}: $v_t = 5860\ m\,s^{-1}$, $\mathbf{e_{t1}}\perp \mathbf{e_l}$, with components $e_{t1}=\{-cos\theta\,cos\phi$, $-cos\theta\,sin\phi$, $sin\theta\}$; and $\mathbf{e_{t2}}\perp\mathbf{e_{t1}}\perp\mathbf{e_l}$, with components $e_{t2}=\{sin\phi$, $-cos\phi$, $0\}$,
 $\rho{=}2330\,kg\,m^{-3}$.\cite{hsueh2014spin} Relaxation time follows $T_1^{-1}\propto B^5$. $B=1\,T$ out-of-plane magnetic field is used. The dependence of $T_1/T_\pi$ and $T_1$(inset) on qubit geometry comes from $k_{ge},\,t,\,\tbkt{\overline{E}}{\hat{q}\cdot r}{\overline{G}}$ terms.

\section{$1/f$ noise}
$1/f$ noise is Gaussian for semiconductors and can be described by the correlation function $S(t)=\langle V(0)V(t)\rangle$. The Fourier transform of this noise correlation function varies inversely proportional to frequency, $S(\omega)=\frac{Ak_BT}{\omega}$, with $A\,{=}\,0.1\,\mu eV$ is estimated from experiments.\cite{takeda2013characterization} The time evolution of spin component is given by,
\begin{eqnarray}
    S_x(t)&=& \langle\langle\sigma_+(t)\rangle\rangle S_{0x}\nonumber\\
    &=& S_{0x}e^{\left(-\int_\infty^\infty d\omega \widetilde{S}_V(\omega)\frac{1}{2}\frac{sin^2(\omega t/2)}{(\omega/2)^2}\right)}\nonumber\\
    &=& S_{0x} e^{-\chi(t)}
\end{eqnarray}
 Using the noise correlation function in the qubit subspace, we can write:
 \begin{equation}
     \chi(t)=\frac{2
     Ak_BT}{\hbar^2}\left(\frac{8k_{ge}^2\epsilon \zeta}{\delta\varepsilon^4}\right)^2 \int_\infty^\infty d\omega \frac{sin^2(\omega t/2)}{\omega^3}
 \end{equation}
 The main contribution to $1/f$ noise comes from the low-frequencies, so a low frequency cut-off $\omega_0$ is chosen as the inverse of measurement time. Assuming, $\omega_0t\ll 1$, and approximating $\int_\infty^\infty d\omega \frac{sin^2(\omega t/2)}{\omega^3}\approx \frac{t^2}{4}\log(\frac{1}{\omega_0t})$, we can write $\chi(t)=(\frac{t}{T_2^*})^2 \log(\frac{1}{\omega_0t})$. This produces:
\begin{equation}\label{eqn:oneoverf}
\left(\frac{1}{T_2^*}\right)_{1/f} \simeq \sqrt{\frac{Ak_BT}{2\hbar^2}}\frac{8 k_{ge}^2\epsilon \zeta}{\delta\varepsilon^4}
\end{equation}
\begin{figure*}[t]
\subfloat[]{\begin{minipage}[c]{0.33\linewidth}
\includegraphics[width= \textwidth]{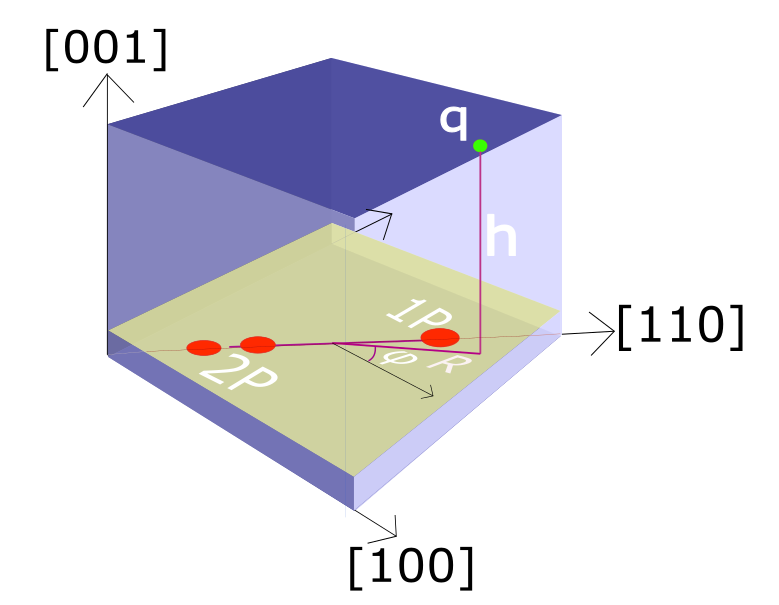}
\end{minipage}}
\hfill
\subfloat[]{\begin{minipage}[c]{0.33\linewidth}
\includegraphics[width= \textwidth]{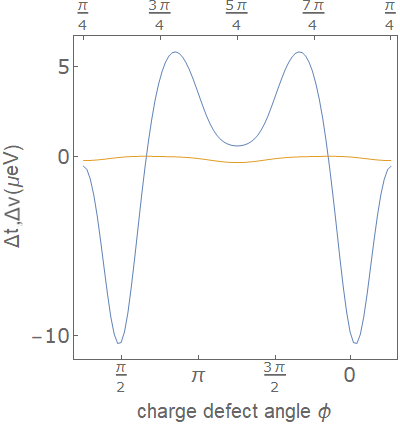}
\end{minipage}}
\hfill
\subfloat[]{\begin{minipage}[c]{0.33\linewidth}
\includegraphics[width= \textwidth]{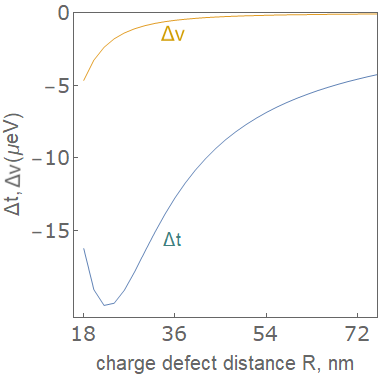}
\end{minipage}}
\hfill
\subfloat[]{\begin{minipage}[c]{0.33\linewidth}
\includegraphics[width=\textwidth]{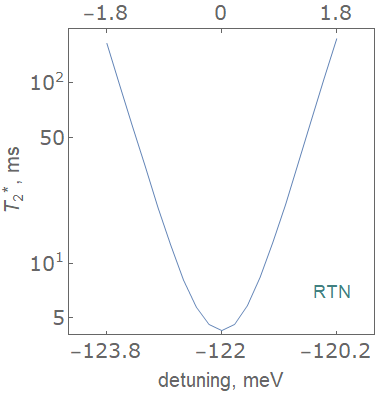}
\end{minipage}}
\hfill
\subfloat[]{\begin{minipage}[c]{0.33\linewidth}
\includegraphics[width=\textwidth]{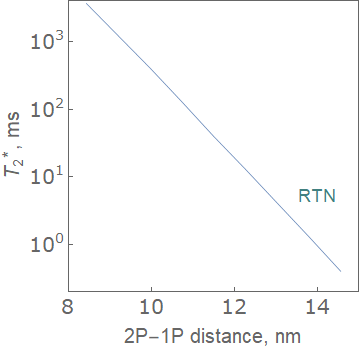}
\end{minipage}}
\hfill
\subfloat[]{\begin{minipage}[c]{0.33\linewidth}
\includegraphics[width=\textwidth]{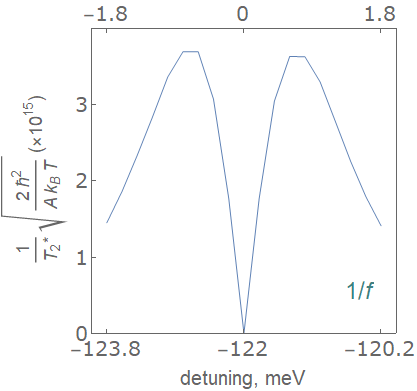}
\end{minipage}}
 \caption{\textbf{Optimising the $2P:1P\parallel [110]$ qubit fidelity in the presence of charge noise}. a)a schematic of a charge defect $q$ placed at an angle $\phi$ w.r.t. to the 2P:1P qubit axis and a distance $R\,$nm away from the center of the qubit i.e. the midpoint between $2P$ and $1P$ dots where the multi donor quantum dot qubit is oriented along $[110]$. For the calculations presented in panel \textit{b-d,f}, the $2P-1P$ distance is $13\,$nm and the $2P-1P$ tunneling energy is $t=1\,$meV. b) variation of fluctuations in the tunneling energy $\Delta t$ and fluctuations in dc detuning bias voltage $\Delta v$ as a function of angular orientation $\phi$ of the charge defect, $R=40\,nm$. The oscillation of $\Delta t$ determines in which directions charge noise should be avoided; for all orientations, $\Delta v$ is nearly constant, and small compared to $\Delta t$. c) Variation of $\Delta t$ and $\Delta v$ as a function of charge defect distance $R$ from the center of the qubit, $\phi =85^o$. d) dephasing time due to random telegraph noise (RTN) $T_{2,RTN}^*$ as a function of detuning ($\delta$), with an external magnetic field of $1 T$. The top label shows the detuning field range to be $\pm2$ meV around the anticrossing. e) $T_{2,RTN}^*$ as a function of $2P-1P$ distance $d$ at anticrossing where the range of $d$ is limited by the perturbation theory. f) The dimensionless part of eqn.~\ref{eqn:oneoverf} for inverse dephasing time $1/T_{2,1/f}^*$, as a function of detuning $\delta$.} 
 \label{fig:noise110}
 \end{figure*}
 \bibliography{reference}